\documentclass[prb,twocolumn,floatfix,superscriptaddress]{revtex4}

\usepackage{amssymb,amsmath,graphicx,afterpage}
\usepackage{color,bm}
\usepackage{cancel}
\pdfoutput=1

\usepackage{color}

\usepackage[colorlinks,bookmarks=false,citecolor=blue,linkcolor=blue,urlcolor=blue]{hyperref}

\newcommand{\colred}[1]{{\color{red}#1}}

\begin{document}

\title{\boldmath Magnetoelectric memory function with optical readout \unboldmath}

\author{V. Kocsis}
\affiliation{RIKEN Center for Emergent Matter Science (CEMS), Wako,
Saitama 351-0198, Japan}
\affiliation{Department of Physics, Budapest University of
Technology and Economics and MTA-BME Lend\"ulet Magneto-optical
Spectroscopy Research Group, 1111 Budapest, Hungary}

\author{K. Penc}
\affiliation{Department of Physics, Budapest University of
Technology and Economics and MTA-BME Lend\"ulet Magneto-optical
Spectroscopy Research Group, 1111 Budapest, Hungary}
\affiliation{Institute for Solid State Physics and Optics, Wigner
Research Centre for Physics, Hungarian Academy of Sciences, H-1525
Budapest, P.O.B. 49, Hungary}

\author{T. R{\~o}{\~o}m}
\affiliation{National Institute of Chemical Physics and Biophysics,
12618 Tallinn, Estonia}

\author{U. Nagel}
\affiliation{National Institute of Chemical Physics and Biophysics,
12618 Tallinn, Estonia}

\author{J. V\'it}
\affiliation{Department of Physics, Budapest University of
Technology and Economics and MTA-BME Lend\"ulet Magneto-optical
Spectroscopy Research Group, 1111 Budapest, Hungary}
\affiliation{Institute of Physics ASCR, Na Slovance 2, 182 21 Prague
8, Czech Republic} \affiliation{Faculty of Nuclear Science and
Physical Engineering, Czech Technical University,
B$\check{r}$ehov\'a 7, 115 19 Prague 1, Czech Republic}

\author{J. Romh\'anyi}
\affiliation{Okinawa Institute of Science and Technology Graduate
University, Onna-son, Okinawa 904-0395, Japan}

\author{Y. Tokunaga}
\affiliation{RIKEN Center for Emergent Matter Science (CEMS), Wako, Saitama 351-0198, Japan}
\affiliation{Department of Advanced Materials Science, University of Tokyo, Kashiwa 277-8561, Japan}

\author{Y. Taguchi}
\affiliation{RIKEN Center for Emergent Matter Science (CEMS), Wako,
Saitama 351-0198, Japan}

\author{Y. Tokura}
\affiliation{RIKEN Center for Emergent Matter Science (CEMS), Wako, Saitama 351-0198, Japan}
\affiliation{Quantum-Phase Electronics Center, Department of Applied Physics, University of Tokyo, Tokyo 113-8656, Japan}
\affiliation{Department of Applied Physics, University of Tokyo, Hongo, Tokyo 113-8656, Japan}

\author{I. K\'ezsm\'arki}
\affiliation{Department of Physics, Budapest University of
Technology and Economics and MTA-BME Lend\"ulet Magneto-optical
Spectroscopy Research Group, 1111 Budapest, Hungary}
\affiliation{Experimental Physics 5, Center for Electronic
Correlations and Magnetism, Institute of Physics, University of
Augsburg, 86159 Augsburg, Germany}

\author{S. Bord\'acs}
\affiliation{Department of Physics, Budapest University of
Technology and Economics and MTA-BME Lend\"ulet Magneto-optical
Spectroscopy Research Group, 1111 Budapest, Hungary}
\affiliation{Hungarian Academy of Sciences, Premium Postdoctor
Program, 1051 Budapest, Hungary}

\maketitle

\textbf{The ultimate goal of multiferroic research is the
development of new-generation non-volatile memory
devices\cite{Fiebig2016}, the so-called magnetoelectric (ME)
memories, where magnetic bits are controlled via electric fields
without the application of electrical currents subject to
dissipation. This low-power operation exploits the entanglement of
the magnetization and the electric polarization coexisting in
multiferroic materials\cite{Kimura2007,Dong2015}. Here we
demonstrate the optical readout of ME memory states in the
antiferromagnetic (AFM) and antiferroelectric (AFE) LiCoPO$_4$,
based on the strong absorption difference of THz radiation between
its two types of ME domains. This unusual contrast is attributed to
the dynamic ME effect of the spin-wave excitations, as confirmed by
our microscopic model, which also captures the characteristics of
the observed static ME effect. Our proof-of-principle study,
demonstrating the control and the optical readout of ME domains in
LiCoPO$_4$, lays down the foundation for future ME memory devices
based on antiferroelectric-antiferromagnetic insulators.}

During the last decades the great potential of multiferroic
materials in realizing ME memory devices has led to the revival of
the ME
effect\cite{Kimura2007,Fiebig2005,Fiebig2005_2,Eerenstein2006,Cheong2007}
and the discovery of a plethora of multiferroic compounds including
BiFeO$_3$, a well characterized room-temperature multiferroic
material\cite{Henron2014,Sando2013,Kezsmarki2015}. In
multiferroics-based memory devices, the writing and reading of
magnetic bits by electric field may be realized via the ME coupling
between the ferromagnetic and ferroelectric orders. Despite the
recent progress, the synthesis of multiferroics with magnetization
and ME effect sufficiently large for applications is still
challenging. As an alternative approach, investigated here,
information could be stored in ME domains even in the absence of
ferromagnetism or ferroelectricity. While a similar concept has been
proposed for metallic compounds, termed as AFM
spintronics\cite{Jungwirth2016}, the potential of AFE-AFM insulators
in ME memories has not been exploited yet. LiCoPO$_4$, being such a
multi-antiferroic insulator, drew attention owing to its strong
linear ME effect\cite{Mercier1967,Rivera1994} and its toroidic
order\cite{VanAken2007,Zimmermann2014}. Here we demonstrate that in
the AFM-AFE phase of LiCoPO$_4$ the two different ME memory states
have distinct optical properties distinguishable by transmission
measurements without the need of high-intensity light
beams\cite{VanAken2007,Zimmermann2014}.

\begin{figure}[h!]
    \begin{center}
    \includegraphics[width=8truecm]{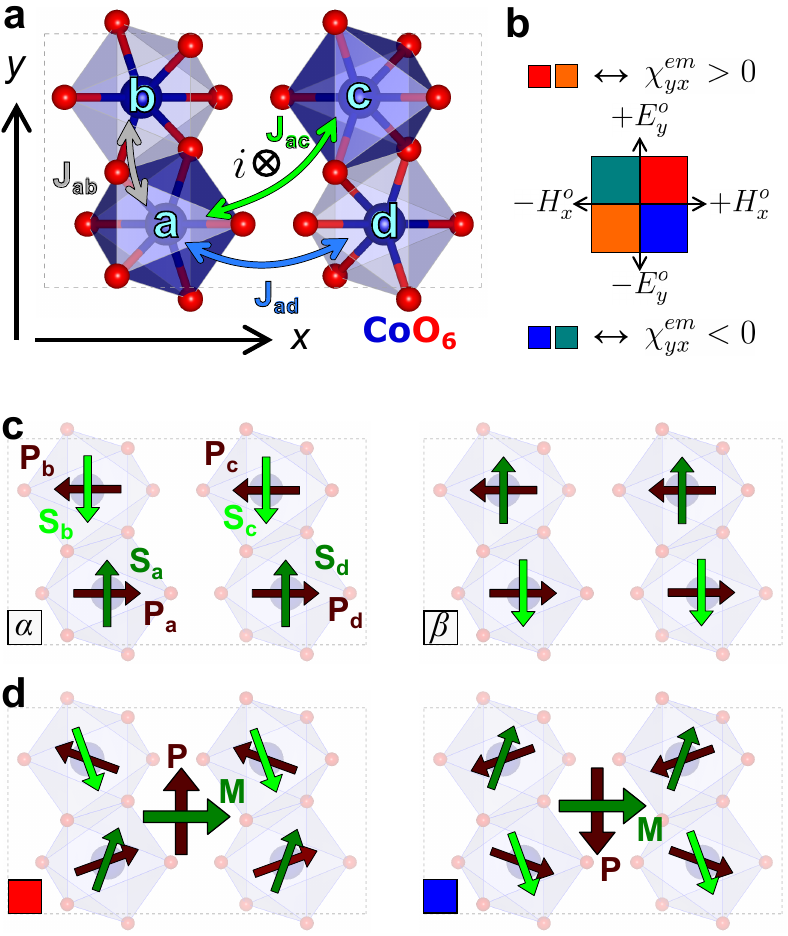}
    \caption{\textbf{$\mid$ Magnetoelectric domains in LiCoPO$_4$.}
    \textbf{a,} Unit cell of the LiCoPO$_4$ viewed from the $z$ axis. The four Co sites (a-d) are surrounded by oxygen octahedra, while Li and P sites are omitted for clarity. The inversion center of the unit cell is labeled by $i$. The three non-equivalent exchange interactions, $J_{ab}$, $J_{ac}$ and $J_{ad}$, are indicated with arrows.
    \textbf{b,} The four combinations $(++), (+-), (-+)$ and $(--)$ of poling fields $(H_x, E_y)$ are represented by four colours.
    \textbf{c,} The magnetic sublattices (green and olive arrows) in the AFM domains $\alpha$ and $\beta$ are interchanged while the polarization pattern (brown arrows) is the same for the two domains.
    \textbf{d,} Domains $\alpha$ and $\beta$ are selected by the poling fields $(++)$ (red) and $(+-)$ (blue) via the ME effect according to
Eq.~\ref{eq:chi_em} assuming $c_{2xy}>0$.}
    \label{lcpo01}
    \end{center}
    \end{figure}

At room temperature LiCoPO$_4$ has the orthorhombic olivine
structure (space group: $Pnma$), which is shown in Fig.~\ref{lcpo01}a.
While each Co site carries a local electric polarization due to its
low site symmetry, the total polarization of the unit cell vanishes
(see Fig.~\ref{lcpo01}c). Below $T_{\rm N}$=21.7\,K, this structural
antiferroelectricity is supplemented by a two-sublattice collinear
AFM order, where $S$=3/2 spins of Co$^{2+}$ ions are aligned parallel
to the $y$ axis\cite{Santoro1966}. Since the AFM state
simultaneously breaks the spatial inversion and the time reversal
symmetries, the material exhibits a linear ME effect ($P_\mu = \chi^{em}_{\mu\nu}H_\nu$, $\mu,\nu=x,y,z$) with finite $\chi^{em}_{xy}$ and
$\chi^{em}_{yx}$ ME susceptibilities\cite{Rivera1994}. Although a
tiny uniform canting of the spins from the $y$ axis may further
reduce the magnetic symmetry and generate finite $\chi^{em}_{xz}$
and $\chi^{em}_{zx}$, these secondary effects are not relevant to
the present study\cite{VanAken2007,Zimmermann2014}.

\begin{figure*}[t!]
    \begin{center}
    \includegraphics[width=17.8truecm]{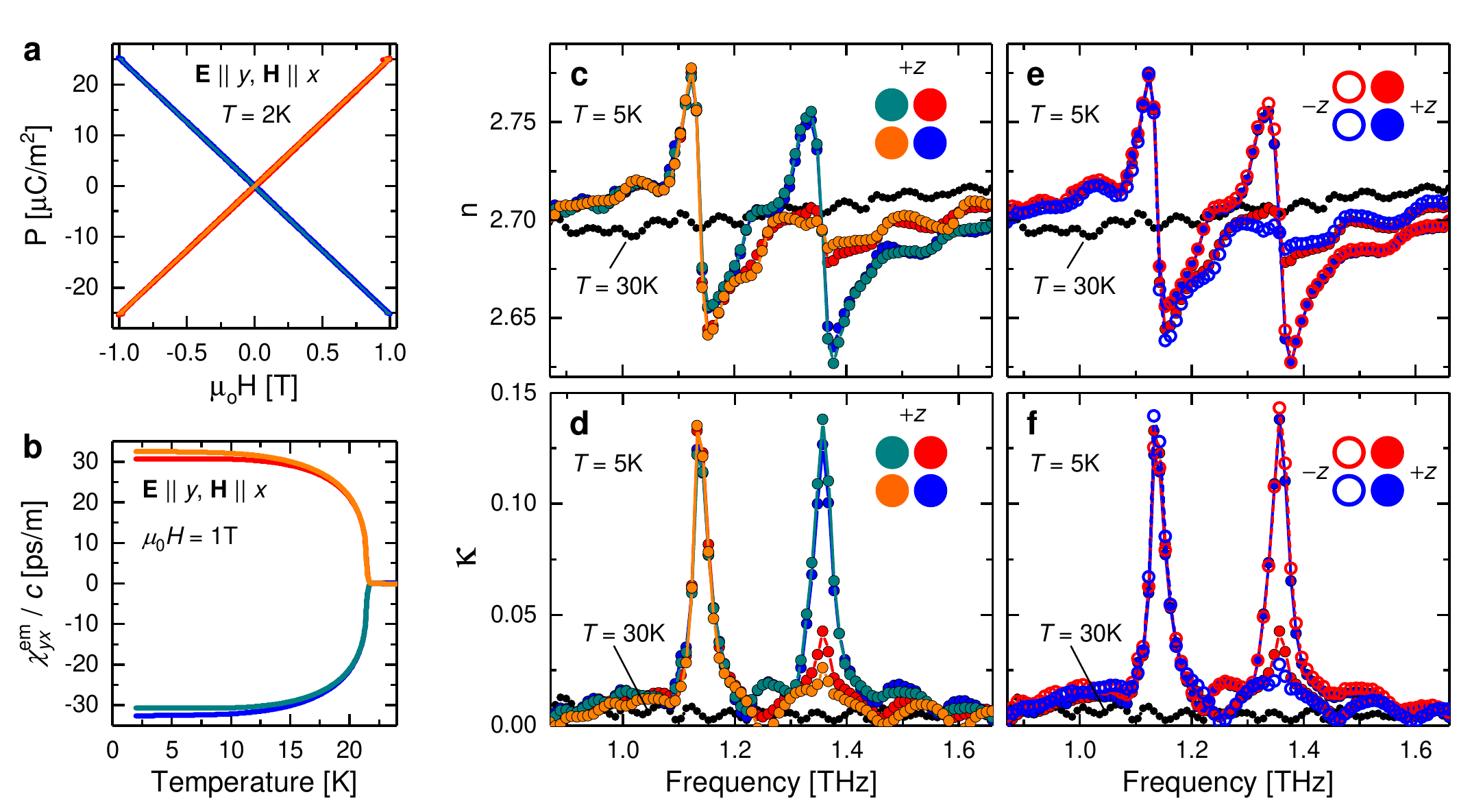}
    \caption{\textbf{$\mid$ Remanent static and optical ME effects in LiCoPO$_4$.}
    \textbf{a,} Magnetic field dependence of the static ME effect at $T$=2\,K measured after poling the sample in the four combinations $(++), (+-), (-+)$ and $(--)$ of poling fields $(H_x, E_y)$. The poling fields were switched off during the measurement, hence the slope of the polarization ($P$) versus magnetic field curve corresponds to the linear ME effect.
    \textbf{b,} Temperature dependence of the linear ME effect, $\chi^{em}_{yx}$, measured in warming up after poling in the four configurations of $(H_x, E_y)$. The colour of each curve in panels \textbf{a} and \textbf{b} corresponds to the applied poling process following the convention introduced in Fig.~\ref{lcpo01}.
    \textbf{c/d,} Spectra of the real/imaginary part of the refractive index at $T$=5\,K measured after poling.
    \textbf{e/f,} Spectra of the real/imaginary part of the refractive index measured at $T$=5\,K after poling in two selected configurations, $(++)$ and $(+-)$. In this case the measurements were performed for light propagation along the $+z$ direction (full symbols) and the $-z$ direction (open symbols). Note that the reversal of the propagation direction is equivalent to the interchange of the two ME domains via the poling process. In panels \textbf{c-f} all spectra were measured using linearly polarized light with $(E^\omega_y,H^\omega_x)$ and spectra measured in the paramagnetic state are plotted in black.}\label{lcpo02}
    \end{center}
    \end{figure*}

In the AFM state two possible domains can exist, labeled as $\alpha$
and $\beta$ in Fig.~\ref{lcpo01}c. These two ME domains can be
transformed into each other by either the spatial inversion or the
time reversal operations, thus, they are characterized by static ME
coefficients $\chi^{em}_{yx}$ of opposite signs, as experimentally
demonstrated in Figs.~\ref{lcpo02}a and b, in agreement with former
studies\cite{Mercier1967,Rivera1994}. Owing to the ME coupling,
simultaneous application of weak crossed fields $E_y
\approx$0.1--1\,kV/cm and $\mu_0H_x$\,$\approx$0.1\,T during the
cooling process through $T_{\rm N}$ establishes the single-domain state.
When the sign of either the electric or the magnetic field is
reversed the other ME domain is selected (see Figs.~\ref{lcpo01}b
and d).

The static ME effect is usually associated with collective modes,
the so-called ME resonances\cite{Kezsmarki2011,Miyahara2012}. These
transitions can be excited by both the electric and magnetic fields
of light as the magnetic component of the radiation generates not
only magnetization but also polarization waves in the material.
Depending on the sign of the optical ME effect, the magnetically
induced polarization waves can interfere either constructively or
destructively with the polarization waves induced by the electric
field of light through the dielectric permittivity, giving rise to
an enhancement or reduction of the complex refractive index
($N=n+\rm{i}\kappa$). For linearly polarized light with
$(E^{\omega}_y, H^{\omega}_x)$ propagating along the $+z$ direction $N_{+z}(\omega)=\sqrt{\varepsilon_{yy}(\omega)\mu_{xx}(\omega)}\pm\chi^{em}_{yx}(\omega)$
,where $\varepsilon_{yy}$ and $\mu_{xx}$ are elements of the
dielectric permittivity and magnetic permeability tensors and $\pm$
signs correspond to the two domains with opposite signs of
$\chi^{em}_{yx}(\omega)$\cite{Kezsmarki2014}. If the optical ME
effect is strong, the ME domain characterized by
$\chi^{em}_{yx}(\omega)<$\,0 can become transparent, while for the
other domain the absorption coefficient,
$\alpha=2\left(\omega/c\right)\kappa$, is enhanced. Such unidirectional
light transmission, also called directional optical anisotropy, has
been reported in several
multiferroics\cite{Kezsmarki2011,Bordacs2012,Kezsmarki2014,Arima2008,Takahashi2012}.
However, this phenomenon has usually been observed in strong
magnetic fields and never as a remanent optical memory effect in
zero field. It is important to note that the contrast between the
two ME domains has to change sign if light propagation direction is
reversed from $+z$ to $-z$ according to $N_{-z}(\omega)=\sqrt{\varepsilon_{yy}(\omega)\mu_{xx}(\omega)}\mp\chi^{em}_{yx}(\omega)$.
Thus, the reversal of the light propagation is expected to be
equivalent with the interchange of the two domains.

Figures~\ref{lcpo02}c-f show the real and imaginary parts of the
refractive index spectra of LiCoPO$_4$ in the terahertz frequency
range for linearly polarized light with $(E^{\omega}_y,
H^{\omega}_x)$. Spectra plotted in Figs.~\ref{lcpo02}c-{d} with four
different colours were obtained after poling the sample from $T>T_{\rm N}$ to $T=$5\,K using four combinations of the poling fields
($\pm H_x$, $\pm E_y$), as described for the static ME measurements.
To observe the remanent effects, the fields were switched off during
the spectroscopic measurements. Below $T_{\rm N}$ two strong
resonances of magnetic origin appear at 1.13\,THz and 1.36\,THz. The
strength of the resonance at 1.36\,THz strongly depends on the
poling conditions, namely it is weak for the same signs and strong
for the opposite signs of poling fields. Moreover, the two spectra
obtained for the same sign of poling fields are identical within the
precision of the experiment as well as the two spectra measured with
poling fields of opposite signs. This indicates the strong ME
character of the mode at 1.36\,THz and also demonstrates the
realization of either of the two ME domain states after the poling
process. In contrast, the mode at 1.13\,THz shows only a weak
optical ME effect, with opposite sign with respect to the strong
effect observed for the mode at 1.36\,THz.

Next, we verified that the optical contrast between the two ME
domains changes upon the reversal of light propagation direction as
expected on symmetry grounds. Indeed, as discerned in
Figs.~\ref{lcpo02}e-{f}, spectra measured for light propagation
along the $+z$ direction with the same sign of poling fields
coincide with spectra measured for light propagation along the $-z$
direction with opposite signs of the poling fields and vice versa.
Due to the optical ME effect for a given direction of light
propagation one of the ME domains is nearly transparent at around
1.36\,THz, while the other domain strongly absorbs photons in this
frequency range, as reflected by the large difference in $\kappa$.

\begin{figure}[t!]
    \begin{center}
    \includegraphics[width=6truecm]{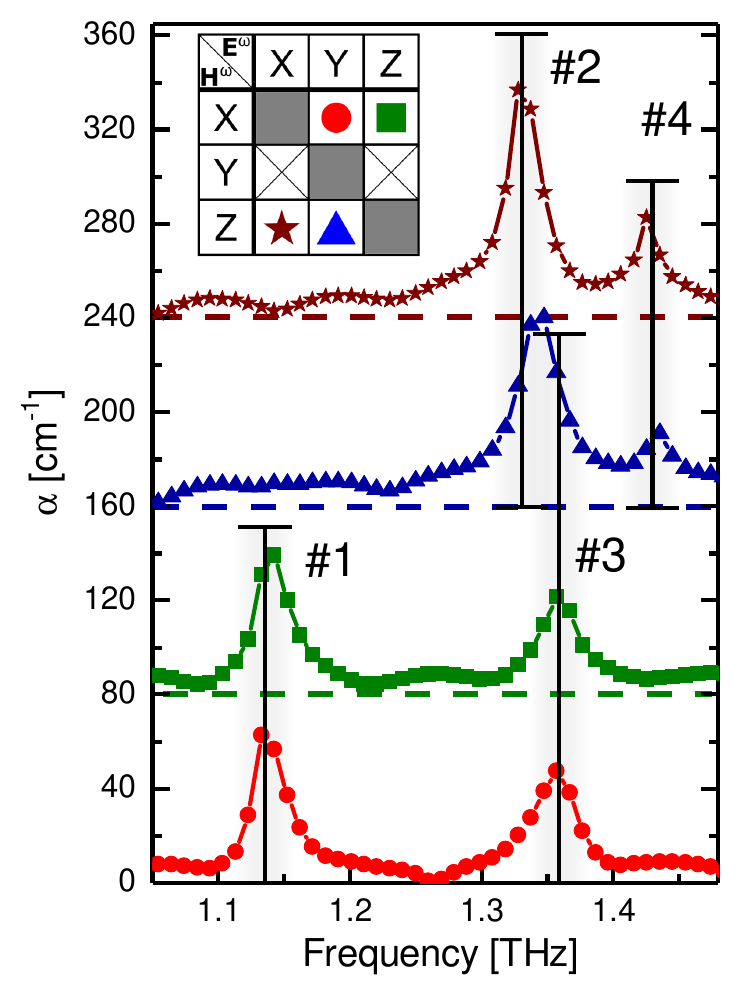}
    \caption{\textbf{$\mid$ Selection rules of the spin-wave excitations in LiCoPO$_4$.}
    Absorption coefficient spectra, $\alpha(\omega)$=$\left(2\omega/c\right)\kappa(\omega)$, measured in six different polarization configurations. The table of the inset indicates the direction of electric ($\mathbf{E^{\omega}}$) and magnetic ($\mathbf{H^{\omega}}$) fields of linearly polarized light. In two polarization configurations with $\mathbf{H^{\omega}}\parallel y$ (not displayed here), no absorption peak was observed. In the remaining four spectra, shifted vertically for clarity, four distinct resonances are identified and labeled as modes \#1 to \#4. The black vertical bars, indicating the positions of these resonances, cross only those spectra where the corresponding resonances are active. The red spectrum, corresponding to the case where the optical ME effect was observed, see Figs.~\ref{lcpo02}c-f, is an average of four different poling combinations.}\label{lcpo03}
    \end{center}
    \end{figure}

In order to systematically determine the selection rules for the two
spin-wave modes observed in Fig.~\ref{lcpo02} and to check the
existence of other spin-wave excitations, optical absorption spectra
were measured for light propagation along the $x$, $y$ and $z$ axes,
with two orthogonal linear polarizations in each case. In the
absence of poling, averaging over the different ME domains
eliminates the directional optical anisotropy term from the
refractive index, hence, $N(\omega)=\sqrt{\varepsilon_{\mu\mu}(\omega)\mu_{\nu\nu}(\omega)}$. As shown in Fig.~\ref{lcpo03}, besides the two modes
coupled to $H^{\omega}_x$ (\#1 and \#3) we observed two additional
spin-wave resonances coupled to $H^{\omega}_z$ at 1.33\,THz (\#2)
and 1.43\,THz (\#4), while no resonance was detected for
$H^{\omega}_y$. The directional optical anisotropy, found to be
strong for mode \#3 and weak for mode \#1 (Figs.~\ref{lcpo02}c-f),
requires that these resonances respond to both $E^{\omega}_y$ and
$H^{\omega}_x$. Indeed, the contribution of mode \#3 to the ($E^\omega_y$,$H^\omega_z$) spectrum (blue in Fig.~\ref{lcpo03}) can only be explained by the electric
dipole excitation of this resonance via $E^{\omega}_y$.

To uncover the mechanism responsible for the static ME effect and
the remanent optical directional anisotropy, we consider the
following Hamiltonian for the four spins ($S_a$, $S_b$, $S_c$,
$S_d$) in the unit cell, imposed by the space group symmetry of
LiCoPO$_4$ (see the Supplementary
Information)\cite{Tian2008,Miyahara2011,Penc2012}:
\begin{align}
\mathcal{H} = & - 4 J_{ij} \sum_{\left\langle{ij}\right\rangle}
\mathbf{S}_i \cdot \mathbf{S}_j - \Lambda_{y^2} \sum_{i}
\left(S^y_{i}\right)^2
-\Lambda_{x^2-z^2} \sum_{i} Q_i^{x^2-z^2}
\nonumber\\
& -\Lambda_{2xz} \left( Q_a^{2xz} - Q_b^{2xz} + Q_c^{2xz} -
Q_d^{xz}\right)
\nonumber\\
& - {g}_{xx} \mu_B H_x \sum_{i} S^x_{i} - E_y P_y.
\label{eq:ham_inv_sites}
\end{align}
where $i\in\{a,b,c,d\}$. $J_{ij}$ stands for the nearest neighbour
exchange coupling with the symmetry-dictated form of
$J_{ab}=J_{cd}$, $J_{ac}=J_{bd}$ and $J_{ad}=J_{cb}$, as indicated
in Fig.~\ref{lcpo01}a. $\Lambda_{y^2}$, $\Lambda_{x^2-z^2}$ and
$\Lambda_{2xz}$ are the single-ion anisotropy parameters and the
spin-quadrupole terms are defined as
$Q_i^{x^2-z^2}=S^x_{i}S^x_{i}-S^z_{i}S^z_{i}$ and
$Q_i^{2\mu\nu}=S^{\mu}_{i}S^{\nu}_{i}+S^{\nu}_{i}S^{\mu}_{i}$. The
last line of Eq.~\ref{eq:ham_inv_sites} describes the interaction
with static magnetic and electric fields, where the electric dipole moment
is calculated following Ref.~\onlinecite{Arima2007}:
\begin{align}
P_y & =
c_{2xy} \left(Q^{2xy}_{a} - Q^{2xy}_{b} - Q^{2xy}_{c} + Q^{2xy}_{d} \right) \nonumber\\
& \phantom{=} + c_{2yz}\left( Q^{2yz}_{a} + Q^{2yz}_{b} -
Q^{2yz}_{c} - Q^{2yz}_{d} \right).\label{eq:pol}
\end{align}

    \begin{figure}[t!]
    \begin{center}
    \includegraphics[width=8.5truecm]{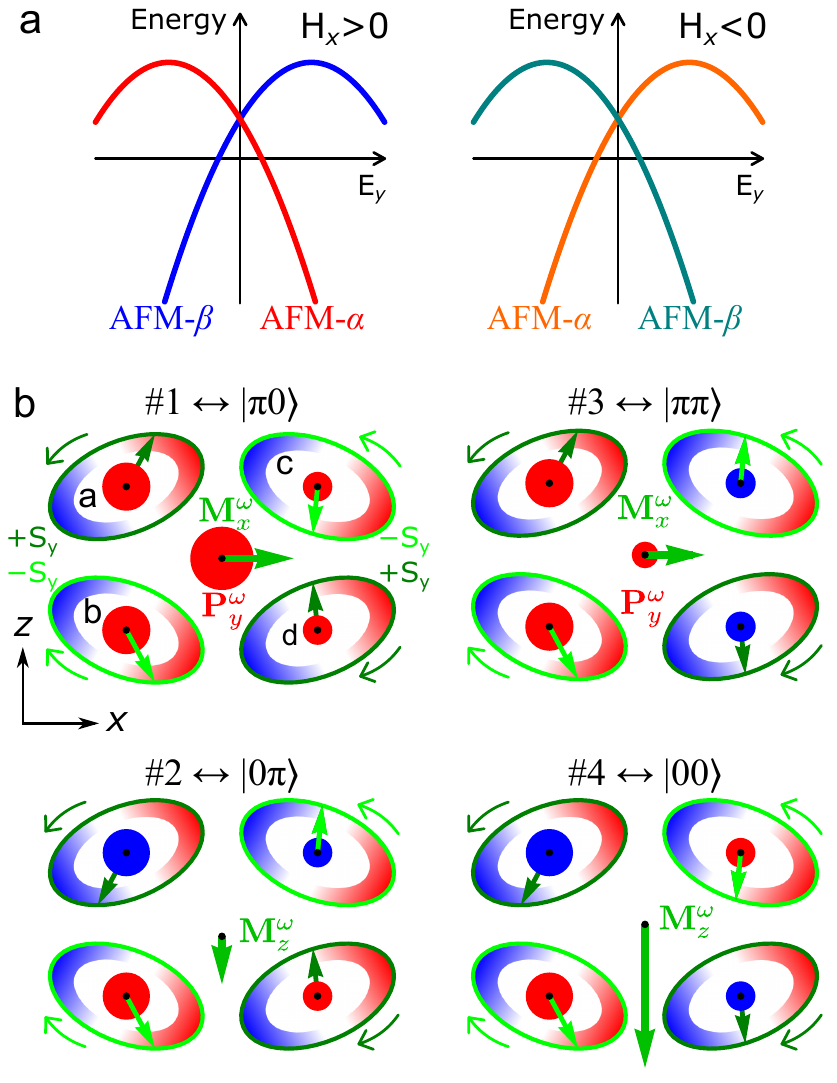}
    \caption{\textbf{$\mid$ Selection of the AFM domains and the spin-excitations of LiCoPO$_4$.}
\textbf{a,} Energies of the AFM domains are quadratic in the
electric and magnetic fields according to Eq.~\ref{eq:energy}. When
$H_x>0$ the $\alpha$ domain has lower energy than the $\beta$ for
positive $E_y$, while negative $E_y$ stabilizes the $\beta$ domain.
For $H_x<0$ role of the two AFM domains are interchanged.
    \textbf{b,} ME ($P^{\omega}_{y}$, $M^{\omega}_{x}$) resonances and magnetic only ($M^{\omega}_{z}$) spin excitations viewed from the $y$ axis, as illustrated on the $\alpha$ domain. Local magnetization of the a and c sites precess counter-clockwise along alternately rotated ellipses in the $xz$ plane, while on the b and d sites spins precess clockwise. The red and blue shading around the ellipses represents the $y$ component of the local polarization, while the green edge the $y$ component of the local magnetization. In the middle of each ellipse the actual direction of the precessing spin is shown by green arrows, while the red and blue marks represent the actual value of the spin-induced polarization. When the precessing spin points to the red (blue) region, the polarization is pointing in the $+y$ ($-y$) axis, while magnitude of the polarization is illustrated by the size of the mark. The resultant oscillating net magnetic ($M^{\omega}_{x}$ and $M^{\omega}_{z}$) and net electric ($P^{\omega}_{y}$) dipole moments are shown in the middle of each unit cell. For $\beta$ domains red and blue shading of the ellipses are reversed, hence $P^{\omega}_{y}$ is in anti-phase compared to the $\alpha$ domain.}\label{lcpo04}
    \end{center}
    \end{figure}

A finite $H_x$ cants the ordered spins, and the non-zero $S_y$ and
$S_x$ components produce a finite electric polarization $P_y$ whose
sign depends on the domain, as schematically shown in
Fig.~\ref{lcpo01}d. The ground state energies of the two AFM domains
are calculated using a variational approach described in the
Supplementary Information:
\begin{align}
E_{GS}(\rm \alpha/\beta) \approx & -18 (J_{ab}+J_{ac}-J_{ad})-9 \Lambda_{y^2}\nonumber\\
&- \frac{3 (2 c_{2xy} E_y\pm{g}_{xx} \mu_B  H_x)^2}{2 (6 J_{ab}+6 J_{ac}+
\Lambda_{y^2})},\label{eq:energy}
\end{align}
where $\pm$ signs correspond to domain $\alpha$ and $\beta$,
respectively. As shown in Fig.~\ref{lcpo04}a, in crossed electric
and magnetic fields, the degeneracy of the two AFM domains is lifted
and $\alpha$ is selected when $E_yH_x>0$, while $\beta$ for
$E_yH_x<0$. The ME susceptibility derived for domain $\alpha$ and
$\beta$ has opposite sign:
\begin{equation}
\chi^{em}_{yx}(\alpha/\beta) = \pm  \frac{6 c_{2xy} g_{xx} \mu_B}{6
J_{ab}+6 J_{ac}+ \Lambda_{y^2}}, \label{eq:chi_em}
\end{equation}
as illustrated in Fig.~\ref{lcpo01}d. This is in accordance with the
experimental observations in Figs.~\ref{lcpo02}a and b.

The oscillating magnetization ($\mathbf{M}^{\omega}$) and
polarization ($\mathbf{P}^{\omega}$) of the spin excitations with
$\Delta{S}=1$ over the ground state were characterized by multiboson
spin-wave theory, which is described in the Supplementary
Information. In agreement with the results of our THz spectroscopy
experiments, two ME excitations were found with $M^{\omega}_x$ and
$P^{\omega}_y$, from which $\vert{\pi0}\rangle$ is assigned to mode
\#1 and $\vert{\pi\pi}\rangle$ to mode \#3. Two further modes,
$\vert{0\pi}\rangle$ and $\vert00\rangle$, are excited with
$H^{\omega}_z$. They are associated with no finite
$\mathbf{P}^{\omega}$ and are assigned to modes \#2 and \#4,
respectively. Motion of the sublattice magnetizations and local
polarizations according to Eq.~\ref{eq:pol} are illustrated for the
$\alpha$ domain in Figs.~\ref{lcpo04}b. The finite
$\mathbf{P}^{\omega}$ of the ME excitations is attributed to the
uncompensated polarization of the unit cell, whereas the local
dynamic polarization is canceled for the $\vert{0\pi}\rangle$ and
$\vert00\rangle$ modes within the $yz$ layers. While the spin
components precess in the same direction in $\alpha$ and $\beta$
domains, there is a $\pi$ phase shift between oscillations of
$P^{\omega}_y$ in the two domains, as Eq.~\ref{eq:pol} is linear in
the sublattice magnetization along the $y$ axis. This sign change of
the dynamic polarization is the microscopic origin of the optical
directional anisotropy in LiCoPO$_4$.

In summary, we have demonstrated that the ME effect can be
exploited for the optical readout of information stored in AFM
domains as the $\pm{k}$ directional optical anisotropy between the
two types of domains gives rise to a sizeable absorption difference
even in the absence of external fields. Main advantages of such type
of memories are i) the possibility to electrically write magnetic
bits with low power consumption via the static ME effect, ii) the
robustness of such devices against stray electric and magnetic
fields due to the dual antiferroic nature of the applied materials,
and iii) the contactless readout function, if the optical scheme
proposed above can be implemented with sufficiently high spatial
resolution.

\section*{Methods}
Single crystals of LiCoPO$_4$ were grown by the optical floating zone method described in Ref.~\onlinecite{SaintMartin2008}.
Plate-shaped samples with 4$\times$4$\times$0.6\,mm$^3$ dimensions were cut for the static and optical measurements.
Measurement of the static ME effect was carried out in a Physical Property Measurement System (Quantum Design) using a Keithley 6517A Electrometer.
Temperature dependence of the ME susceptibility was calculated from the polarization measured in the warming runs in the presence of 1\,T magnetic field.
Time-domain THz spectroscopy was used to measure the complex
refractive index spectra in the 200\,GHz - 2\,THz frequency range.
The THz radiation was guided by off-axis parabolic mirrors, and its
precise linear polarization was maintained by free standing wire grid
polarizers, placed into parallel THz beam before and after the
sample. THz light generation was based on a Toptica Teraflash
spectrometer\cite{Vieweg2014} whose fs light pulses were coupled to
the emitter and receiver photoconductive antennas by optical fibers.
This arrangement provided an easy way to reverse the propagation
direction of the THz radiation by interchanging the position of the
emitter and receiver, while leaving the optical path intact.
Optical measurements with reversed light propagation were done when the sample was cooled to a single ME domain state.
In order to align the ME domains of LiCoPO$_4$ electric field in the
range of 0.1--1\,kV/cm and the magnetic field 0.1\,T of a permanent
magnet were applied along the $y$ and $x$ axes, respectively, at
$T$=30\,K, above $T_{\rm N}$.
In the next step, the sample was cooled down to $T$=5\,K, where the poling fields were switched off and then the transmission measurements were carried out.
THz absorption experiments using a Martin-Puplett interferometer in NICPB, Tallinn were used to find suitable $E_y$ and $H_x$ fields for poling.

\textbf{Acknowledgements} This work was supported by the Hungarian
Research Funds OTKA K 108918, OTKA PD 111756, OTKA K106047, National
Research, Development and Innovation Office  NKFIH, ANN 122879 and
Bolyai 00565/14/11, by the Deutsche Forschungsgemeinschaft (DFG) via
the Transregional Research Collaboration TRR 80: From Electronic
Correlations to Functionality (Augsburg - Munich - Stuttgart) and by
the Estonian Ministry of Education and Research under Grant No.
IUT23-03, and the European Regional Development Fund project TK134.

\textbf{Author Contributions} V.K., S.B., J.V., T.R., U.N. performed
the measurements; V.K., S.B., I.K., J.V. analysed the data; V.K.,
Y.Tokunaga prepared the sample; K.P., J.R. developed the theory;
V.K., K.P., I.K. wrote the manuscript; each author contributed to
the discussion of the results; V.K, Y. Taguchi, S.B., I.K. planned and
supervised the project.

\textbf{Additional information} The authors declare no competing financial interests.

\cleardoublepage

\section*{Supplementary material}

\begin{figure}[htbp]
\begin{center}
\includegraphics[width=8truecm]{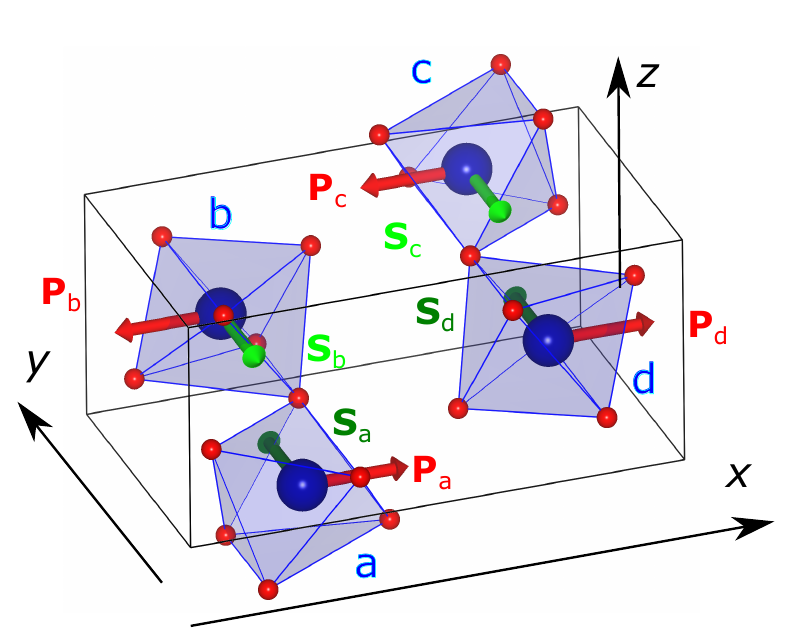}
\caption{\textbf{$\mid$ Unit cell of LiCoPO$_4$ exemplified on the $\alpha$ AFM domain.} The sites a together with b, and c together with d form separate layers, which are connected by inversion symmetry. Spin orientation are labeled by green arrows, dark green arrows encode $S^y>0$, while light green spins are $S^y<0$. Local polarizations are shown by red arrows. For the $\alpha$ ME domain the cross product points to the $+z$, while for the $\beta$ to the $-z$ direction.}
\label{supplfig01} 
\end{center}
\end{figure}

\begin{table}[t]
\begin{tabular}{cccccccc}
E & $C^{(z)}_2$ & $C^{(y)}_2$ & $C^{(x)}_2$ & $i$ & $\sigma_{xy}$ & $\sigma_{xz}$ & $\sigma_{yz}$\\
\noalign{\smallskip}
\hline
\hline
\noalign{\smallskip}
a & b & c & d & c & d & a & b\\
\noalign{\smallskip}
b & a & d & c & d & c & b & a\\
\noalign{\smallskip}
c & d & a & b & a & b & c & d\\
\noalign{\smallskip}
d & c & b & a & b & a & d & c\\
\noalign{\smallskip}
\hline
\noalign{\smallskip}
$x$ & $-x+\frac{1}{2}$ & $-x$ & $x+\frac{1}{2}$ & $-x$ & $x+\frac{1}{2}$ & $x$ & $-x+\frac{1}{2}$ \\
\noalign{\smallskip}
$y$ & $-y$ & $y+\frac{1}{2}$ & $-y+\frac{1}{2}$ & $-y$ & $y$ & $-y+\frac{1}{2}$ & $y+\frac{1}{2}$\\
\noalign{\smallskip}
$z$ & $z+\frac{1}{2}$ & $-z$ & $-z+\frac{1}{2}$ & $-z$ & $-z+\frac{1}{2}$ & $z$ & $z+\frac{1}{2}$\\
\noalign{\smallskip}
\hline
\noalign{\smallskip}
$P_x$ & $-P_x$ & $-P_x$ & $P_x$ & $-P_x$ & $P_x$ & $P_x$ & $-P_x$ \\
\noalign{\smallskip}
$P_y$ & $-P_y$ & $P_y$ & $-P_y$ & $-P_y$ & $P_y$ & $-P_y$ & $P_y$\\
\noalign{\smallskip}
$P_z$ & $P_z$ & $-P_z$ & $-P_z$ & $-P_z$ & $-P_z$ & $P_z$ & $P_z$\\
\noalign{\smallskip}
\hline
$S^x$ & $-S^x$ & $-S^x$ & $S^x$ & $S^x$ & $-S^x$ & $-S^x$ & $S^x$\\
\noalign{\smallskip}
$S^y$ & $-S^y$ & $S^y$ & $-S^y$ & $S^y$ & $-S^y$ & $S^y$ & $-S^y$\\
\noalign{\smallskip}
$S^z$ & $S^z$ & $-S^z$ & $-S^z$ & $S^z$ & $S^z$ & $-S^z$ & $-S^z$\\
\noalign{\smallskip}
\hline
$Q^{2xy}_a$ & $Q^{2xy}_b$ & $-Q^{2xy}_c$ & $-Q^{2xy}_d$ & $Q^{2xy}_c$ & $Q^{2xy}_d$ & $-Q^{2xy}_a$ & $-Q^{2xy}_b$\\
\noalign{\smallskip}
$Q^{2yz}_a$ & $-Q^{2yz}_b$ & $-Q^{2yz}_c$ & $Q^{2yz}_d$ & $Q^{2yz}_c$ & $-Q^{2yz}_d$ & $-Q^{2yz}_a$ & $Q^{2yz}_b$\\
\noalign{\smallskip}
$Q^{2xz}_a$ & $-Q^{2xz}_b$ & $Q^{2xz}_c$ & $-Q^{2xz}_d$ & $Q^{2xz}_c$ & $-Q^{2xz}_d$ & $Q^{2xz}_a$ & $-Q^{2xz}_b$\\
\noalign{\smallskip}
\hline
\end{tabular}
\caption{\textbf{$\mid$ Effect of  the $Pnma$  space group on coordinates, magnetic moments, spin-multipoles and Co sites.}  The components of the electric polarization behave as the corresponding $x$, $y$, and $z$ coordinates factorized by the fractional displacement. As an example, from the transformation properties of the spin-quadrupolar $Q^{2\mu\nu}=S^\mu S^\nu+S^\nu S^\mu$ operator we can read off the symmetry allowed single-ion anisotropies in the Hamiltonian: $\Lambda_{2xz,a}=-\Lambda_{2xz,b}=-\Lambda_{2xz,c}=\Lambda_{2xz,d}=\Lambda_{2xz}$, while $\Lambda_{2xy,i}=0$ and $\Lambda_{2yz,i}=0$ for all $i=$a,b,c,d.
}\label{tab:tablePnma} 
\end{table}

\begin{table}[t]
\begin{tabular}{cccccccc}
E & $C^{(z)}_2$ & \colred{$C'^{(y)}_2$} & \colred{$C'^{(x)}_2$} & \colred{$i'$} & \colred{$\sigma'_{xy}$} & $\sigma_{xz}$ & $\sigma_{yz}$\\
\noalign{\smallskip}
\hline
\hline
\noalign{\smallskip}
$P_x$ & $-P_x$ & $-P_x$ & $P_x$ & $-P_x$ & $P_x$ & $P_x$ & $-P_x$ \\
\noalign{\smallskip}
$P_y$ & $-P_y$ & $P_y$ & $-P_y$ & $-P_y$ & $P_y$ & $-P_y$ & $P_y$\\
\noalign{\smallskip}
$P_z$ & $P_z$ & $-P_z$ & $-P_z$ & $-P_z$ & $-P_z$ & $P_z$ & $P_z$\\
\noalign{\smallskip}
\hline
$S^x$ & $-S^x$ & \colred{$S^x$} & \colred{$-S^x$} & \colred{$-S^x$} & \colred{$S^x$} & $-S^x$ & $S^x$\\
\noalign{\smallskip}
$S^y$ & $-S^y$ & \colred{$-S^y$} & \colred{$S^y$} & \colred{$-S^y$} & \colred{$S^y$} & $S^y$ & $-S^y$\\
\noalign{\smallskip}
$S^z$ & $S^z$ & \colred{$S^z$} & \colred{$S^z$} & \colred{$-S^z$} & \colred{$-S^z$} & $-S^z$ & $-S^z$\\
\noalign{\smallskip}
\hline
\end{tabular}
\caption{\textbf{$\mid$ Effect of the $Pnma'$ magnetic space group on the magnetic moments of Co$^{2+}$ ions.} Symmetry elements combined with time reversal symmetry in the $Pnma'$ space group are indicated by red color. Spin-quadrupoles have the same transformation properties under $Pnma'$ as under the paramagnetic space group in Table~\ref{tab:tablePnma}. This is the symmetry group of the time-reversal broken ground state, described in Eq.~(\ref{eq:S_inv_Pnma'}).
\label{tab:tablePnma'}
} 
\end{table}

\section{Symmetry analysis}
 
Below $T_{\rm N}=21.7$\,K the magnetic moments of LiCoPO$_4$ order antiferromagnetically, with the moments parallel to the $y$-axis\cite{Santoro1966} as shown in Fig.~\ref{supplfig01}. 
Symmetry of the crystal in the paramagnetic phase is the $Pnma$, while in the magnetically ordered phase the $Pnma'$ magnetic space group, elements of which are enumerated in Table~\ref{tab:tablePnma} and Table~\ref{tab:tablePnma'}, respectively. 
The spatial inversion symmetry prevents the development of finite polarization in the unit cell. 
However, besides the AFM order, the magnetic space group of LiCoPO$_4$ allows antiferroelectric (AFE) order of the local polarization.
The magnetic structure in the ground state is given by
\begin{align}
\mathbf{S}_a = -\mathbf{S}_b = -\mathbf{S}_c =\mathbf{S}_d = \left(
\begin{array}{c}
 0 \\ \mu \\ 0
\end{array}
 \right),
 \label{eq:S_inv_Pnma'}
\end{align}
while the symmetry allowed local electric dipole moments at the Co sites are
\begin{align}
\mathbf{P}_a = -\mathbf{P}_c = \left(
\begin{array}{c}
 \xi \\ 0 \\ \zeta
\end{array}
 \right),
\mathbf{P}_b = -\mathbf{P}_d = \left(
\begin{array}{c}
 -\xi \\ 0 \\ \zeta
\end{array}
 \right).
 \label{eq:P_inv_Pnma'}
\end{align}
The local polarization has two independent AFE components along the $x$ and $z$ directions. In Fig.~\ref{supplfig01} we show only the component along the $x$ axis for the sake of simplicity. The local polarization may have different origins; one is inherent to the distorted CoO$_6$ clusters, while the other is due to the spins via the $p-d$ hybridization model. In the model presented below we concentrate only on the latter case, which will give rise to the magnetoelectric effect. 

Transformation properties of the spin-multipolar moments and permutation of the Co sites under the $Pnma$ space group impose restrictions to the possible terms in the minimal spin Hamiltonian (Table~\ref{tab:tablePnma}).
As a result, the minimal Hamiltonian for the four spins in the unit cell, assuming periodic boundary conditions is \begin{align}
\mathcal{H} & = \phantom{+}
4 J_{ab} (\mathbf{S}_a\cdot\mathbf{S}_b + \mathbf{S}_c\cdot\mathbf{S}_d) 
+4 J_{ac} (\mathbf{S}_a\cdot\mathbf{S}_c + \mathbf{S}_b\cdot\mathbf{S}_d)
\nonumber\\ & \phantom{=}
+ 4 J_{ad} (\mathbf{S}_a\cdot\mathbf{S}_d + \mathbf{S}_b\cdot\mathbf{S}_c) 
\nonumber\\
& - \Lambda_{y^2} \left[\left(S^y_{a}\right)^2 +\left(S^y_{b}\right)^2 +\left(S^y_{c}\right)^2 +\left(S^y_{d}\right)^2\right] 
\nonumber\\
&-\Lambda_{x^2-z^2} \left(
Q^{x^2-z^2}_{a} + Q^{x^2-z^2}_{b} + Q^{x^2-z^2}_{c} + Q^{x^2-z^2}_{d}
\right)
\nonumber\\
&
-\Lambda_{2xz} \left(
Q^{2xz}_{a} - Q^{2xz}_{b} + Q^{2xz}_{c} - Q^{2xz}_{d} \right),
\label{eq:ham_inv_sites}
\end{align}
The interaction between the spins is described by the isotropic Heisenberg exchanges with the $J_{ab}$, $J_{ac}$ and $J_{ad}$ coupling contants. The exchange anisotropies (as the Dzyaloshinskii-Moriya interaction and symmetric exchange anisotropies) are disregarded here as they are assumed to be weak, and the magnetic anisotropies in LiCoPO$_4$ are taken care of by the $\Lambda$ single-ion anisotropies. 
As shown in Table~\ref{tab:tablePnma}, the Hamiltonian is invariant under the $Pnma$ space group if  $\Lambda_{2xy}=0$ and  $\Lambda_{2yz}=0$.
The remaining single-ion anisotropies, with coefficients $\Lambda_{y^2}$, $\Lambda_{x^2-z^2}$, and $\Lambda_{2xz}$ describe an anisotropy tensor with a principal axis along the $y$ direction and two axes in the $xz$ plane.
Throughout this paper we will assume that the easy axis magnetic anisotropy of LiCoPO$_4$ is dominated by the $\Lambda_{y^2}>0$ parameter, {\it i.e.} $\Lambda_{y^2}\gg \Lambda_{x^2-z^2},\Lambda_{2xz}$.
As a further simplification we also introduce the notation for spin-quadrupoles:
\begin{subequations}
\begin{align}
 Q^{x^2-z^2}_{i} &= S^x_{i} S^x_{i} - S^z_{i} S^z_{i} \;,
\\
 Q^{2\mu\nu}_{i} &= S^\mu_{i} S^\nu_{i} + S^\nu_{i} S^\mu_{i} \;,
\end{align}
\end{subequations}
where $\mu,\nu=x,y,z$ and $i=$\,a,b,c,d. Strictly speaking, there are five spin-quadrupolar operators ($Q^{3y^2-S^2}$, $Q^{x^2-z^2}$ and $ Q^{2\mu\nu}$), however, we decided to replace $Q^{3y^2-S^2}$ by $\left(S^y\right)^2$ as it differs from the commonly used definition for the on-site easy-axis anisotropy.

From symmetry we get the following expressions for the magnetizations 
\begin{subequations}
\begin{align}
M_x & = g_{xx} \left( S^x_a+S^x_b+S^x_c+S^x_d \right) 
\nonumber\\&\phantom{=}
+ g_{xz} \left( S^z_a-S^z_b+S^z_c-S^z_d \right), 
\label{eq:Mxdefs}
 \\
M_y & = g_{yy} \left( S^y_a+S^y_b+S^y_c+S^y_d \right), 
 \\
M_z & = g_{zz} \left( S^z_a+S^z_b+S^z_c+S^z_d \right), 
\nonumber\\
&\phantom{=}
 + g_{xz} h_z \left( S^x_a-S^x_b+S^x_c-S^x_d \right).
\end{align}
\label{eq:Mdefs}
\end{subequations}
and for the electric polarizations:
\begin{subequations}
\begin{align}
P_x &= 
b_{y^2} \left[(S^y_a)^2-(S^y_b)^2-(S^y_c)^2+(S^y_d)^2\right] 
\nonumber\\
&+b_{x^2-z^2} 
\left(
Q^{x^2-z^2}_{a} -Q^{x^2-z^2}_{b}-Q^{x^2-z^2}_{c} + Q^{x^2-z^2}_{d}
\right) 
\nonumber\\
&+ b_{2xz} \left(Q^{2xz}_{a} + Q^{2xz}_{b} - Q^{2xz}_{c} - Q^{2xz}_{d}\right), 
\label{eq:Px}
\\
P_y & = 
c_{2xy} \left(Q^{2xy}_{a} - Q^{2xy}_{b} - Q^{2xy}_{c} + Q^{2xy}_{d} \right) \nonumber\\
& \phantom{=}
+ c_{2yz}\left( Q^{2yz}_{a} + Q^{2yz}_{b} - Q^{2yz}_{c} - Q^{2yz}_{d} \right),
\label{eq:Py}
\\
P_z &= 
d_{y^2} \left[(S^y_a)^2+(S^y_b)^2-(S^y_c)^2-(S^y_d)^2\right] 
\nonumber\\
&+ d_{x^2-z^2} 
\left(
Q^{x^2-z^2}_{a} + Q^{x^2-z^2}_{b} - Q^{x^2-z^2}_{c} - Q^{x^2-z^2}_{d}
\right) 
\nonumber\\
&+ d_{2xz} \left(Q^{2xz}_{a} - Q^{2xz}_{b} - Q^{2xz}_{c} + Q^{2xz}_{d}\right).
\label{eq:Pz}
\end{align}
\label{eq:Pdefs}
\end{subequations}
Here we note, that although the Hamiltonian can contain combination of $Q^{x^2-z^2}$, $Q^{2xz}$ and $(S^y)^2$, it cannot have neither $Q^{2xy}$ nor $Q^{2yz}$ elements.
Moreover, it is also not possible to express the Hamiltonian in terms of $P_\nu$ ($\nu=x,y,z$).
Nevertheless, the local $P_y$ at each site transforms as the $Q^{2xy}$ and $Q^{2yz}$ operators (c.f. Table.~\ref{tab:tablePnma}), therefore it can be represented by the linear combination of these spin-quadrupolar operators.
Both the static and dynamic ME effects are expressed by the couplings between the operators and the corresponding physical quantities; $c_{2xy}$ and $c_{2yz}$ for $\chi^{em}_{xy}$.

Interactions with the external magnetic and electric fields are described by
\begin{subequations}
\begin{align}
\mathcal{H}_\text{Zeeman} &= - H_x M_x - H_y M_y - H_z M_z 
\,, \\
\mathcal{H}_{EP} &= - E_x P_x - E_y P_y - E_z P_z\,.
\end{align}
\end{subequations}

\section{Variational treatment (mean field)}

To describe the static properties of LiCoPO$_4$ at low temperatures, we will treat our model using a site-factorized wave function as a variational Ansatz for the ground state:
\begin{align}
 | \Psi^{\text{GS}}_{\text{var}}\rangle = 
 | \Psi_{1,a}\rangle 
 | \Psi_{1,b}\rangle 
 | \Psi_{1,c}\rangle 
 | \Psi_{1,d}\rangle
\end{align}
We shall minimize the 
\begin{align}
 E_{\text{var}}=\frac{\langle \Psi^{\text{GS}}_{\text{var}}| \mathcal{H}| \Psi^{\text{GS}}_{\text{var}}\rangle}{\langle \Psi^{\text{GS}}_{\text{var}}| \Psi^{\text{GS}}_{\text{var}}\rangle}
\end{align}
variational energy, by optimizing the wave functions on the sites, $| \Psi_{1,a}\rangle$\dots$| \Psi_{1,d}\rangle$. The variational setup is similar to the case of Ba$_2$CoGe$_2$O$_7$, therefore we implemented the procedure applied there\cite{Romhanyi2011,Penc2012,Romhanyi2012}.
 First, we will consider the problem in the absence of the external fields ($\mathbf{H}=0$ and $\mathbf{E}=0$). After this, we will turn on the fields to describe the effect of poling. 

It is convenient to work in a basis where the quantization axis is along the $y$ direction, 
\begin{subequations}
\begin{align}
|\Uparrow_y \rangle &= \frac{1}{\sqrt{8}} \left(|\Uparrow\rangle +  i \sqrt{3} |\uparrow\rangle -\sqrt{3} |\downarrow\rangle -i |\Downarrow\rangle\right)
\\
|\uparrow_y \rangle &= \frac{1}{\sqrt{8}} \left(\sqrt{3}|\Uparrow\rangle + i  |\uparrow\rangle + |\downarrow\rangle + i\sqrt{3} |\Downarrow\rangle\right)
\\
|\downarrow_y \rangle &= \frac{1}{\sqrt{8}} \left(\sqrt{3}|\Uparrow\rangle - i  |\uparrow\rangle + |\downarrow\rangle - i\sqrt{3} |\Downarrow\rangle\right)
\\
|\Downarrow_y \rangle &= \frac{1}{\sqrt{8}} \left(|\Uparrow\rangle -  i \sqrt{3} |\uparrow\rangle -\sqrt{3} |\downarrow\rangle + i |\Downarrow\rangle\right)
\end{align}
\end{subequations}
so that the $|\Uparrow_y \rangle$, $|\uparrow_y \rangle$, $|\downarrow_y \rangle$, and $|\Downarrow_y \rangle$ are the eigenfunctions of the $ S^y = \tilde S^0 $ operator with eigenvalues $3/2$, $1/2$, $-1/2$, and $-3/2$, respectively.  
The off-diagonal spin operators in the rotated frame are 
\begin{align}
 S^z &= \frac{\tilde S^+ + \tilde S^-}{2} \quad\text{and}\quad
 S^x = \frac{\tilde S^+ - \tilde S^-}{2i}.
\end{align}

 The minimum is achieved with the
\begin{subequations}
\begin{align}
 |\Psi_{1,a}\rangle = & |\Uparrow_y \rangle - \sqrt{3} \gamma |\downarrow_y\rangle \;, \\
 |\Psi_{1,b}\rangle = & |\Downarrow_y\rangle - \sqrt{3} \gamma |\uparrow_y\rangle \;, \\
 |\Psi_{1,c}\rangle = & |\Downarrow_y\rangle - \sqrt{3} \bar\gamma |\uparrow_y\rangle \;,\\
 |\Psi_{1,d}\rangle = & |\Uparrow_y \rangle - \sqrt{3} \bar\gamma |\downarrow_y\rangle \;,
\end{align}
\label{eq:psiGS_h0}
\end{subequations}
site-dependent wave functions with energy
\begin{align}
E_{o}^{\text{var}} &=-18 (J_{ab}+J_{ac}-J_{ad})-9 \Lambda_{y^2} 
\nonumber\\& \phantom{+}
-\frac{6 \left(\Lambda_{x^2-z^2}^2+ \Lambda_{2xz}^2\right)}{6 J_{ab}+6 J_{ac}-6 J_{ad}+ \Lambda_{y^2} } \;.
\end{align} 
In Eqs.~(\ref{eq:psiGS_h0}) the $\gamma$ is a complex number determined by the parameters of the exchange field and the on--site anisotropies,
\begin{align}
\gamma = \frac{ \lambda_{x^2-z^2} - i \lambda_{2xz}}{2} \;, 
\label{eq:gammadef}
\end{align}
with 
\begin{subequations}
\begin{align}
 \lambda_{x^2-z^2} &= \frac{ \Lambda_{x^2-z^2} }{6 J_{ab} + 6 J_{ac} - 6 J_{ad} + \Lambda_{y^2}},
\\
 \lambda_{2xz} &= \frac{ \Lambda_{2xz} }{6 J_{ab} + 6 J_{ac} - 6 J_{ad} + \Lambda_{y^2}}.
\end{align}
\end{subequations}

The expectation values of the $S^x$ and $S^z$ are zero on all four sites, only the $S^y$ matrix elements are nonzero:
\begin{align}
\frac{\langle \Psi_{1,a} |
S^y_a| \Psi_{1,a}\rangle}{\langle \Psi_{1,a} | \Psi_{1,a}\rangle}
& \approx
\frac{3}{2}-6 \gamma \bar\gamma +\cdots
\end{align}
and the other sites follow the AFM pattern given by Eq.~(\ref{eq:S_inv_Pnma'}) for a proper choice of the $J$ exchange couplings.

Notably,  due to the single-ion anisotropies the wave function describes spins, length of which is shorter than 3/2.
On the other hand the spin acquires quadrupolar features, as exemplified by the expectation values of the spin-quadrupolar operators, e.g. on site a:
 \begin{subequations}
\begin{align}
\langle \Psi_{1,a} |
 S^y_a S^y_a 
| \Psi_{1,a}\rangle
& \approx
\frac{9}{4}-6 \gamma \bar\gamma
\\
\langle \Psi_{1,a} | Q^{x^2-z^2}_a | \Psi_{1,a}\rangle
& \approx 3 ( \gamma + \bar\gamma) 
\\
\langle \Psi_{1,a} |
Q^{2xz}_{a}
| \Psi_{1,a}\rangle
& \approx 
3 i (\gamma -\bar\gamma)
\\
\langle \Psi_{1,a} |
Q^{2xy}_{a}
| \Psi_{1,a}\rangle 
& = 0
\\
\langle \Psi_{1,a} |
Q^{2yz}_{a}
| \Psi_{1,a}\rangle 
& = 0
\end{align}
\end{subequations}

Here we note also that the the wave functions on the sites $a$ and $c$ are time reversal pairs, and so are the ones on sites $b$ and $d$. In fact, the wave functions of the other AFM ground state are obtained by site permutations $a \leftrightarrow c$ and $b \leftrightarrow d$, as the anisotropies of the local Hamiltonian are the same for $a$ and $c$ sites, only the direction of the local Weiss field is opposite.
Performing the same permutation of the expression for the polarization operators $P_x$ and $P_y$, Eqs.~(\ref{eq:Px}) and (\ref{eq:Py}), their sign changes. This already hints at the interaction between the N\'eel state and the polarizations.

\subsection{Poling with $h_x$ and $E_y$}

The inclusion of external fields into the problem will enlarge the zero field variational wave function given in Eqs.~(\ref{eq:psiGS_h0}) to allow for the canting of the spins, 
\begin{subequations}
\begin{align}
 |\Psi_{1,a}\rangle = & |\Uparrow_y \rangle + i \sqrt{3} \eta |\uparrow_y\rangle - \sqrt{3} \gamma |\downarrow_y\rangle \;, \\
 |\Psi_{1,b}\rangle = & |\Downarrow_y\rangle - i \sqrt{3} \eta |\downarrow_y\rangle - \sqrt{3} \gamma |\uparrow_y\rangle \;, \\
 |\Psi_{1,c}\rangle = & |\Downarrow_y\rangle - i \sqrt{3} \bar\eta |\downarrow_y\rangle- \sqrt{3} \bar\gamma |\uparrow_y\rangle \;,\\
 |\Psi_{1,d}\rangle = & |\Uparrow_y \rangle + i \sqrt{3} \bar\eta |\uparrow_y\rangle - \sqrt{3} \bar\gamma |\downarrow_y\rangle \;,
\end{align}
\end{subequations}
where the energy minimum is achieved by
\begin{align}
 \eta &= \frac{g_{xx} H_x + 2 c_{2xy} E_y }{4(6 J_{ab}+6 J_{ac}+ \Lambda_{y^2})} 
 - i \frac{g_{xz} H_x + 2 c_{2yz} E_y }{4(6 J_{ac}-6 J_{ad}+ \Lambda_{y^2})} \;,
\end{align}
providing the ground state energy in finite external fields,
\begin{align}
E_{\rm GS}(\alpha) &= E_0
- \frac{3 (2 c_{2xy} E_y+g_{xx} H_x)^2}{2 (6 J_{ab}+6 J_{ac}+ \Lambda_{y^2} )}
\nonumber
\\& \phantom{+}
 - \frac{3 (2 c_{2yz} E_y+g_{xz} H_x)^2}{2 (6 J_{ac}-6 J_{ad}+ \Lambda_{y^2} )}.
\end{align}\label{eq:GSenergyalpha}
Canting of the spins on site 'a'  is proportional to the variational parameter $\eta$,
\begin{align} 
\langle \Psi_{1,a} | \mathbf{S} | \Psi_{1,a} \rangle =
\frac{3}{2} 
\left(
\begin{array}{c}
\eta+\bar\eta\\
1\\
i(\eta-\bar\eta)\\
\end{array}
\right) \,.
\end{align}
The  symmetry in finite $E_y$ and $H_x$ is reduced to the $\text{P}\text{m}2^\prime \text{m}^\prime$ magnetic space group with remaining elements $\{ E,C'^{(y)}_2,\sigma'_{xy},\sigma_{yz}\}$ taken from Table~\ref{tab:tablePnma'}. We
note that exactly the same elements are missing for either a finite $E_y$ only, or a finite  $H_x$ only, or when both $E_y$ and  $H_x$ are finite. This is reflected in the variational solution as well, since 
\begin{align}
 \langle S_a^x \rangle &= \langle S_b^x \rangle = \langle S_c^x \rangle = \langle S_d^x \rangle \,,
 \\
 \langle S_a^y \rangle &= -\langle S_b^y \rangle = -\langle S_c^y \rangle = \langle S_d^y \rangle \,,
 \\
 \langle S_a^z \rangle &= -\langle S_b^z \rangle =  \langle S_c^z \rangle = -\langle S_d^z \rangle \,,
\end{align}
as anticipated from the form of the $M_x$, Eq.~(\ref{eq:Mxdefs}). Using Eq.~\ref{eq:GSenergyalpha} the magnetoelectric susceptibility is 
\begin{align}
\chi^{em}_{yx,\alpha} &= - \frac{\partial^2 E}{\partial H_x \partial E_y} 
\nonumber\\
&=\frac{6 c_{2xy} g_{xx}}{6 J_{ab}+6 J_{ac}+ \Lambda_{y^2}}
+ \frac{6 c_{2yz} g_{xz}}{6 J_{ac}-6 J_{ad}+ \Lambda_{y^2} }\,.
\end{align}
We found that the leading term of the magnetoelectric susceptibility is independent from the $\Lambda_{x^2-z^2}$ and $\Lambda_{2xz}$ on-site anisotropies, while the term containing $g_{xz}$ is expected to be a minute correction.

Solution for the other N\'eel AFM domain is given by the
\begin{subequations}
\begin{align}
 |\Psi_{1,a}'\rangle = & |\Downarrow_y\rangle - i \sqrt{3} \bar\eta' |\downarrow_y\rangle - \sqrt{3} \bar\gamma |\uparrow_y\rangle \;, \\
 |\Psi_{1,b}'\rangle = & |\Uparrow_y \rangle + i \sqrt{3} \bar\eta' |\uparrow_y\rangle - \sqrt{3} \bar\gamma |\downarrow_y\rangle \;, \\
 |\Psi_{1,c}'\rangle = & |\Uparrow_y \rangle + i \sqrt{3} \eta' |\uparrow_y\rangle - \sqrt{3} \gamma |\downarrow_y\rangle \;,\\
 |\Psi_{1,d}'\rangle = & |\Downarrow_y\rangle - i \sqrt{3} \eta' |\downarrow_y\rangle- \sqrt{3} \gamma |\uparrow_y\rangle \;.
\end{align}
\end{subequations}
wave functions, with
\begin{align}
 \eta' &= \frac{g_{xx} H_x - 2 c_{2xy} E_y }{4(6 J_{ab}+6 J_{ac}+ \Lambda_{y^2})} 
 - i \frac{g_{xz} H_x - 2 c_{2yz} E_y }{4(6 J_{ac}-6 J_{ad}+ \Lambda_{y^2})} \;.
\end{align}
In this case the spin expectation values on site 'a' are
\begin{align} 
\langle \Psi_{1,a} | \mathbf{S} | \Psi_{1,a} \rangle =
\frac{3}{2} 
\left(
\begin{array}{c}
\eta'+\bar\eta'\\
-1\\
i(\eta'-\bar\eta')\\
\end{array}
\right)
\end{align}
and the energy in finite fields is:
\begin{align}
E_{\rm GS}(\beta) &= E_0
- \frac{3 (2 c_{2xy} E_y-g_{xx} H_x)^2}{2 (6 J_{ab}+6 J_{ac}+ \Lambda_{y^2} )}
\nonumber
\\& \phantom{+}
 - \frac{3 (2 c_{2yz} E_y-g_{xz} H_x)^2}{2 (6 J_{ac}-6 J_{ad}+ \Lambda_{y^2} )}.
\end{align}
The polarizations and the susceptibilities change sign for the two AFM domains ($\alpha$ and $\beta$), i.e.:
\begin{equation}
\chi^{em}_{yx}(\beta) = -\chi^{em}_{yx}(\alpha) \,.
\end{equation}

To merge the solutions achieved for the ME susceptibility of the two AFM domains we may write:
\begin{equation}
\chi^{em}_{yx}(\alpha/\beta) =
\pm\frac{6 c_{2xy} g_{xx}}{6 J_{ab}+6 J_{ac}+ \Lambda_{y^2}}
\pm\frac{6 c_{2yz} g_{xz}}{6 J_{ac}-6 J_{ad}+ \Lambda_{y^2} },
\end{equation}
where the $\pm$ sign holds for the $\alpha$ and $\beta$ domains. 
As the sign of the denominator is expected to be positive, sign of the magnetoelectric susceptibility for the $\alpha$ and $\beta$ is mutually determined by the sign of the material specific $c_{2xy}$ constant.
The ME susceptibility of the the $\alpha$($\beta$) domain can be positive (negative) for $c_{2xy}>0$ and negative (positive) for $c_{2xy}<0$.
This means -- as expected -- that the two domains are interchangeable in the interpretations, although their ME response has opposite sign.
At this point there is no way to determine the sign of the $c_{2xy}$ parameter, therefore we may fix it positive for the sake of simplicity.

\section{Multiboson spin wave}

Below we will use the multiboson spin-wave theory to analyze the excitation spectrum. Since the excited state is created by light, we only need to look at the $\Gamma$ point in the Brillouin zone, keeping in mind that our unit cell contains 4 magnetic ions. Here we closely follow the calculation presented in Refs.~[\onlinecite{Penc2012}] and [\onlinecite{Romhanyi2012}].

 The starting point for the multiboson spin-wave theory is the product form of the ground state wave function, and the bosons are associated with the wave function on a site, which include the ground state and the local excitations. For example, for site 'a' and in the lowest order in the $\Lambda$ on-site anisotropies, the wave functions
\begin{subequations}
\begin{align}
 |\Psi_{1,a}\rangle &= |\Uparrow_y \rangle - \sqrt{3} \gamma |\downarrow_y \rangle
 \\
 |\Psi_{2,a}\rangle &= |\uparrow_y \rangle - \sqrt{3} \gamma' |\Downarrow_y \rangle 
 \label{eq:psi2a}
 \\
 |\Psi_{3,a}\rangle &= |\downarrow_y \rangle + \sqrt{3} \bar\gamma |\Uparrow_y \rangle
 \label{eq:psi3a}
 \\
 |\Psi_{4,a}\rangle &= |\Downarrow_y \rangle + \sqrt{3} \bar \gamma' |\uparrow_y \rangle
\end{align}
\end{subequations}
span a four dimensional Hilbert space, built up by wave functions localised to site 'a'. Here $\gamma$ is defined in Eq.~(\ref{eq:gammadef}) while $\gamma'$ reads
\begin{align}
 \gamma' & = \frac{ \Lambda_{x^2-z^2} - i \Lambda_{2xz} }{12 J_{ab} + 12 J_{ac} - 12 J_{ad} - 2 \Lambda_{y^2}} \;.
\end{align}
Similarly, together with Eq.~(\ref{eq:psi2a}), the first excited states on the four sites are
\begin{subequations}
\begin{align}
 |\Psi_{2,b}\rangle \propto & |\downarrow_y\rangle - \sqrt{3}\gamma' |\Uparrow_y\rangle \;, \\
 |\Psi_{2,c}\rangle \propto & |\downarrow_y\rangle - \sqrt{3}\bar\gamma' |\Uparrow_y\rangle \;, \\
 |\Psi_{2,d}\rangle \propto & |\uparrow_y\rangle - \sqrt{3}\bar\gamma' |\Downarrow_y\rangle \;.
\end{align}
\label{eq:psi2}
\end{subequations}

We keep only the bosons describing the four lowest energy excitations given by the wave functions shown in Eq.~(\ref{eq:psi2}), and we use the following labeling :
\begin{subequations}
\begin{align}
b^{\phantom{\dagger}}_{a}
&\to 
\frac{1}{2} \left(
b^{\phantom{\dagger}}_{00}+b^{\phantom{\dagger}}_{\pi\pi}+b^{\phantom{\dagger}}_{\pi0}+b^{\phantom{\dagger}}_{0\pi}
\right),
\\
b^{\phantom{\dagger}}_{b}
&\to 
\frac{1}{2} \left(
b^{\phantom{\dagger}}_{00}
-b^{\phantom{\dagger}}_{\pi\pi}
-b^{\phantom{\dagger}}_{\pi0}
+b^{\phantom{\dagger}}_{0\pi}
\right),
\\
b^{\phantom{\dagger}}_{c}
&\to 
\frac{1}{2} \left(
b^{\phantom{\dagger}}_{00}
-b^{\phantom{\dagger}}_{\pi\pi}
+b^{\phantom{\dagger}}_{\pi0}
-b^{\phantom{\dagger}}_{0\pi}
\right),
\\
b^{\phantom{\dagger}}_{d}
&\to 
\frac{1}{2} \left(
b^{\phantom{\dagger}}_{00}
+b^{\phantom{\dagger}}_{\pi\pi}
-b^{\phantom{\dagger}}_{\pi0}
-b^{\phantom{\dagger}}_{0\pi}
\right) \;.
\end{align}
\end{subequations}
The spin wave Hamiltonian can be separated into diagonal and off-diagonal parts:
\begin{subequations}
\begin{align}
\mathcal{H} = \mathcal{H}^{\text{diag}}+ \mathcal{H}^{\text{offdiag}} \,.
\end{align}
\end{subequations}
For $\lambda_{2xz}=0$ only the $\mathcal{H}^{\text{diag}}$ exists and takes a block-diagonal form:
\begin{align}
\mathcal{H}^{\text{diag}} = 
\mathcal{H}_{00}^{\text{diag}} 
+ \mathcal{H}_{0\pi}^{\text{diag}}
+ \mathcal{H}_{\pi0}^{\text{diag}}
+ \mathcal{H}_{\pi\pi}^{\text{diag}} \;,
\end{align}
with 
\begin{align}
\mathcal{H}_{q_1q_2}^{\text{diag}} &= 
\Omega_{q_1q_2} b^{\dagger}_{q_1q_2} b^{\phantom{\dagger}}_{q_1q_2} 
\nonumber\\&\phantom{=}
+ \frac{1}{2}\Xi_{q_1q_2} 
( b^{\dagger}_{q_1q_2} b^{\dagger}_{q_1q_2} + b^{\phantom{\dagger}}_{q_1q_2} b^{\phantom{\dagger}}_{q_1q_2} )
 \;,
 \label{eq:Hq1q2diag}
\end{align}
where the $\Omega_{q_1q_2}$ are 
\begin{subequations}
\begin{align}
\Omega_{00} &=  6 J_{ab} + 6 J_{ac} + 2 \Lambda_{y^2} 
\nonumber\\&\phantom{=}
-12 (J_{ab}+J_{ac}) \lambda_{x^2-z^2} \;,
 \label{eq:Omega_00}
 \\
\Omega_{\pi\pi} &=  6 J_{ab} + 6 J_{ac} + 2 \Lambda_{y^2} 
\nonumber\\&\phantom{=}
+ 12 (J_{ab}+J_{ac}) \lambda_{x^2-z^2}  \;,
 \label{eq:Omega_pipi}
 \\
\Omega_{0\pi} &=  6 J_{ab} + 6 J_{ac} - 12 J_{ad} + 2 \Lambda_{y^2} 
\nonumber\\&\phantom{=}
- 12 (J_{ab}-J_{ac}) \lambda_{x^2-z^2}  \;,
 \label{eq:Omega_0pi}
 \\
\Omega_{\pi0} &=  6 J_{ab} + 6 J_{ac} - 12 J_{ad} + 2 \Lambda_{y^2} 
\nonumber\\&\phantom{=}
+ 12 (J_{ab}-J_{ac}) \lambda_{x^2-z^2}  \;,
 \label{eq:Omega_pi0}
\end{align}
\end{subequations}
and the $\Xi_{q_1q_2}$ are
\begin{subequations}
\begin{align}
\Xi_{00} &= 6 J_{ab} + 6 J_{ac} - 12 J_{ad} \lambda_{x^2-z^2} \;,
\\
\Xi_{\pi\pi} &= -6 J_{ab} - 6 J_{ac} - 12 J_{ad} \lambda_{x^2-z^2} \;,
\\
\Xi_{0\pi} &= 6 J_{ab} - 6 J_{ac} + 12 J_{ad} \lambda_{x^2-z^2} \;,
\\
\Xi_{\pi0} &= - 6 J_{ab} + 6 J_{ac} + 12 J_{ad} \lambda_{x^2-z^2} \;.
\end{align}
\end{subequations}
The off-diagonal  part $\mathcal{H}^{\text{offdiag}}$ is proportional to $\lambda_{2xz}$, and  introduces interaction between the different modes of the diagonal Hamiltonian:,
\begin{align}
\mathcal{H}^{\text{offdiag}} 
&= 
 12 i J_{ac} \lambda_{2xz} ( 
  b^{\dagger}_{0\pi} b^{\phantom{\dagger}}_{00} 
- b^{\dagger}_{00} b^{\phantom{\dagger}}_{0\pi} 
)
\nonumber\\
&\phantom{=}
- 12 i J_{ac} \lambda_{2xz} ( 
  b^{\dagger}_{\pi0} b^{\phantom{\dagger}}_{\pi\pi} 
- b^{\dagger}_{\pi\pi} b^{\phantom{\dagger}}_{\pi0} 
)\;.
\label{eq:Hoffdiag}
\end{align}

\subsection{Excitation energies}

First, we consider the case of $\lambda_{2xz}=0$. A Bogoliubov-Valatin transformation provides the eigenvalues of the $\mathcal{H}_{q_1q_2}^{\text{diag}}$ operators [Eq.~(\ref{eq:Hq1q2diag})], as it involves solving $2\times 2$ matrices:
\begin{align}
\omega_{q_1,q_2}
&= \sqrt{\Omega_{q_1,q_2}^2-\Xi_{q_1,q_2}^2} \;.
\end{align}

In the absence of the $\lambda_{x^2-z^2}$ the spin wave energies are  two-fold degenerate, with energies
\begin{subequations}
\begin{align}
\omega_{0,0}
&= \omega_{\pi,\pi}
= 2 \sqrt{
\left(
6 J_{ab} + 6 J_{ac} + \Lambda_{y^2}
\right)
 \Lambda_{y^2} 
} \;,
\label{eq:omega00omegapipi}
\\
\omega_{0,\pi}
&= \omega_{\pi,0}
= 2 \sqrt{
\left(
6 J_{ab} - 6 J_{ad} + \Lambda_{y^2}
\right)
\left(
6 J_{ac} - 6 J_{ad} + \Lambda_{y^2} 
\right) 
} \;.
 \label{eq:omega0piomegapi0}
\end{align}
\end{subequations}
\begin{widetext}
A finite  $\lambda_{x^2-z^2}$ value splits the degeneracy, and the energies are
\begin{subequations}
\begin{align}
\omega_{00} & = 2 \sqrt{ \Lambda_{y^2} (6 J_{ab} +6 J_{ac} + \Lambda_{y^2} )- 12 ( J_{ab} + J_{ac} ) (3 J_{ab} +3 J_{ac} -3 J_{ad} + \Lambda_{y^2} ) \lambda_{x^2-z^2} }
\label{eq:omega00}
\\
\omega_{\pi\pi} & = 2 \sqrt{ \Lambda_{y^2} (6 J_{ab} +6 J_{ac} + \Lambda_{y^2} ) + 12 ( J_{ab} + J_{ac} ) (3 J_{ab} +3 J_{ac} -3 J_{ad} + \Lambda_{y^2} ) \lambda_{x^2-z^2} }
\label{eq:omegapipi}
\\
\omega_{0\pi} & = 2 \sqrt{(6 J_{ab} -6 J_{ad} + \Lambda_{y^2} )(6 J_{ac} -6 J_{ad} + \Lambda_{y^2} )- 12 ( J_{ab} - J_{ac} ) (3 J_{ab} +3 J_{ac} -3 J_{ad} + \Lambda_{y^2} ) \lambda_{x^2-z^2} }
\label{eq:omega0pi}
\\
\omega_{\pi0} & = 2 \sqrt{(6 J_{ab} -6 J_{ad} + \Lambda_{y^2} )(6 J_{ac} -6 J_{ad} + \Lambda_{y^2} ) + 12 ( J_{ab} - J_{ac} ) (3 J_{ab} +3 J_{ac} -3 J_{ad} + \Lambda_{y^2} ) \lambda_{x^2-z^2} }
\label{eq:omegapi0}
\end{align}
\label{eq:omegas}
\end{subequations}

For finite values of the $\lambda_{2xz}$, the problem described by the $\mathcal{H}^{\text{offdiag}}$, given by Eq.~(\ref{eq:Hoffdiag}), becomes equivalent to a $4\times 4$  generalized eigenvalue problem. In order to achieve an analytic solution, we consider the  $\lambda_{2xz}$ as a small parameter, and treat $\mathcal{H}^{\text{offdiag}}$ perturbatively. It turns out that the main consequence of the finite $\lambda_{2xz}$ is the mixing of the eigenvectors of the unperturbed solution, which will effect the transition  matrix elements only. The eigenvalues are changing only as $\lambda_{2xz}^2$, which  can be safely neglected. Therefore, we will keep the same labels ($00$,$0\pi$,$\pi0$,$\pi\pi$) of the unperturbed excitations  for both $\lambda_{2xz}=0$ and finite $\lambda_{2xz}$.

\subsection{The dynamical response}

To address the strength of the absorption of the modes for different polarizations of the light, we need to calculate the imaginary part  of the magnetic and electric susceptibilities. At zero temperature, the  imaginary part of the magnetic susceptibility is given as
\begin{align}
\text{Im} \chi_{\nu\nu}^{mm} = \pi \sum_f \langle  \left|\langle f | M^\nu | \text{GS} \rangle \right|^2
\left[ \delta(\omega-E_f+E_{\text{GS}}) - \delta(\omega+E_f-E_{\text{GS}})\right]\;,
\label{eq:chimm}
\end{align}
where the summation is over the $f$ final states, with energy $E_f$, and $\nu=x,y,z$. A similar expression holds for $\text{Im} \chi_{\nu\nu}^{ee}$, with the  magnetization $M^\nu$ replaced by the $P^\nu$ polarization. 
Strength of the directional optical anisotropy depends on the imaginary part  of the magnetoelectric susceptibility, which at zero temperature reads:
\begin{align}
\text{Im} \chi_{yx}^{em} = \pi \sum_f \langle  \text{GS} | P^y | f \rangle \langle f | M^x | \text{GS} \rangle 
\left[ \delta(\omega-E_f+E_{\text{GS}}) - \delta(\omega+E_f-E_{\text{GS}})\right]\,.
\label{eq:chiem}
\end{align}
Therefore, to calculate the dynamical susceptibilities, we need to express the magnetizations  given by Eqs.~(\ref{eq:Mdefs}) with the bosonic operators. We get
\begin{subequations}
\begin{align}
M^x
&=
\sqrt{3}\left[ g_{xz} - ( g_{xz} \lambda_{x^2-z^2} - g_{xx} \lambda_{2xz} ) \right] 
( b^{\dagger}_{\pi0} + b^{\phantom{\dagger}}_{\pi0} )
+i \sqrt{3} \left[ g_{xx} + ( g_{xx} \lambda_{x^2-z^2} + g_{xz} \lambda_{2xz} ) \right] 
( b^{\dagger}_{\pi\pi} - b^{\phantom{\dagger}}_{\pi\pi} ) ,
\\
M^z
&=
\sqrt{3}\left[ g_{zz} - ( g_{zz} \lambda_{x^2-z^2} - g_{xz} \lambda_{2xz} ) \right]
( b^{\dagger}_{00} + b^{\phantom{\dagger}}_{00} )
+i \sqrt{3} \left[ g_{xz} + ( g_{xz} \lambda_{x^2-z^2} + g_{zz} \lambda_{2xz} ) \right]
( b^{\dagger}_{0\pi} - b^{\phantom{\dagger}}_{0\pi} ) \,,
\end{align}
\end{subequations}
while the $M^y$ has matrix elements with higher energy magnetic excitations, which are disregarded here.
Out of the three polarization operators in  Eqs.~(\ref{eq:Pdefs}), only the $P^y$ couples to the lowest energy magnons:
\begin{align}
P^y
&=
2 \sqrt{3} c_{2yz} 
( b^{\dagger}_{\pi0} + b^{\phantom{\dagger}}_{\pi0} )+2 i \sqrt{3} c_{2xy} 
( b^{\dagger}_{\pi\pi} - b^{\phantom{\dagger}}_{\pi\pi} ).
\end{align}
From the equations above, we can conclude that the $|00\rangle$ and $|0\pi\rangle$ modes are purely magnetic modes, excited with the magnetic field only, and the 
$|\pi\pi\rangle$ and $|\pi0\rangle$ modes are magnetoelectric modes, excited by both the magnetic and electric component of the incident light.
  
  After a tedious calculation, the transition matrix elements for the $M^z$ in the purely magnetic $|00\rangle$ and $|0\pi\rangle$ modes, with the energies given by Eqs.~(\ref{eq:omega00}) and (\ref{eq:omega0pi}), respectively, are
\begin{subequations}
\begin{align}
\langle 00 | 
b^{\dagger}_{00} + b^{\phantom{\dagger}}_{00} 
|\text{GS}\rangle
& = 
\left(\frac{ \Lambda_{y^2} }{6 J_{ab} +6 J_{ac} + \Lambda_{y^2} }\right)^{1/4},
\\
\langle 00 | 
b^{\dagger}_{0\pi} - b^{\phantom{\dagger}}_{0\pi}
|\text{GS}\rangle
& = i \lambda_{2xz} \frac{ J_{ac} (6 J_{ab} +3 J_{ac} -3 J_{ad} + \Lambda_{y^2} )}{ (3 ( J_{ab} - J_{ad} ) ( J_{ac} - J_{ad} )- J_{ad} \Lambda_{y^2} )}
\left(\frac{ \Lambda_{y^2} }{6 J_{ab} + 6 J_{ac} + \Lambda_{y^2} }\right)^{1/4},
\end{align}
\end{subequations}
and
\begin{subequations}
\begin{align}
\langle 0\pi | 
b^{\dagger}_{00} + b^{\phantom{\dagger}}_{00}
|\text{GS}\rangle
& = i \lambda_{2xz} 
\frac{ J_{ac} (3 J_{ac} -3 J_{ad} + \Lambda_{y^2} )}{(3 ( J_{ab} - J_{ad} ) ( J_{ac} - J_{ad} )- J_{ad} \Lambda_{y^2} )}
\left(
\frac{6 J_{ab} -6 J_{ad} + \Lambda_{y^2} }{6 J_{ac} -6 J_{ad} + \Lambda_{y^2} }
\right)^{1/4},
\\
\langle 0\pi | 
b^{\dagger}_{0\pi} - b^{\phantom{\dagger}}_{0\pi}
|\text{GS}\rangle
& =
\left(\frac{6 J_{ab} -6 J_{ad} + \Lambda_{y^2} }{6 J_{ac} -6 J_{ad} + \Lambda_{y^2} }\right)^{1/4},
\end{align}
\end{subequations}
in the leading order in $\lambda_{2xz}$.
The matrix elements in Eq.~(\ref{eq:chimm}) are then
\begin{align}
\langle 00 |M^z |\text{GS}\rangle
& \approx
\sqrt{3} g_{zz} 
\left(\frac{ \Lambda_{y^2} }{6 J_{ab} +6 J_{ac} + \Lambda_{y^2} }\right)^{1/4}
\end{align}
and
\begin{align}
\langle 0\pi |M^z | \text{GS} \rangle
&\approx
i \sqrt{3} \left[ g_{xz} 
+  g_{xx} \lambda_{2xz}
\frac{(J_{ac}-J_{ad}) (3 J_{ab}+3 J_{ac}-3 J_{ad}+ \Lambda_{y^2})}{3 ( J_{ab} - J_{ad} ) ( J_{ac} - J_{ad} ) -J_{ad} \Lambda_{y^2}}
 \right] 
\left(
\frac{6 J_{ab} -6 J_{ad} + \Lambda_{y^2} }{6 J_{ac} -6 J_{ad} + \Lambda_{y^2} }
\right)^{1/4}  
\end{align}
for the $|00\rangle$ and the $|0\pi\rangle$ modes.

The $|\pi0\rangle$ and $|\pi\pi\rangle$ modes  have both finite $M_x$ and $P_y$ transition matrix elements, so these modes show optical directional anisotropy.
The matrix elements for the $|\pi\pi\rangle$ mode, with energy $\omega_{\pi\pi}$, Eq.~(\ref{eq:omegapipi}), are
\begin{subequations}
\begin{align}
\langle \pi\pi | 
b^{\dagger}_{\pi\pi} - b^{\phantom{\dagger}}_{\pi\pi}
|\text{GS}\rangle
& = 
\left(\frac{ \Lambda_{y^2} }{6 J_{ab} +6 J_{ac} + \Lambda_{y^2} }\right)^{1/4} ,
\\
\langle \pi\pi | 
b^{\dagger}_{\pi0} + b^{\phantom{\dagger}}_{\pi0}
|\text{GS}\rangle
& = - \lambda_{2xz} i \frac{ J_{ac} (6 J_{ab} +3 J_{ac} -3 J_{ad} + \Lambda_{y^2} )}{ (3 ( J_{ab} - J_{ad} ) ( J_{ac} - J_{ad} )- J_{ad} \Lambda_{y^2} )}
\left(\frac{ \Lambda_{y^2} }{6 J_{ab} + 6 J_{ac} + \Lambda_{y^2} }\right)^{1/4} ,
\end{align}
\end{subequations}
and for the $|\pi0\rangle$ mode, with energy $\omega_{\pi0}$, Eq.~(\ref{eq:omegapi0}), are
\begin{subequations}
\begin{align}
\langle \pi0 | 
b^{\dagger}_{\pi\pi} - b^{\phantom{\dagger}}_{\pi\pi}
|\text{GS}\rangle
& = -i \lambda_{2xz} 
\frac{ J_{ac} (3 J_{ac} -3 J_{ad} + \Lambda_{y^2} )}{(3 ( J_{ab} - J_{ad} ) ( J_{ac} - J_{ad} )- J_{ad} \Lambda_{y^2} )}
\left(
\frac{6 J_{ab} -6 J_{ad} + \Lambda_{y^2} }{6 J_{ac} -6 J_{ad} + \Lambda_{y^2} }
\right)^{1/4} ,
\\
\langle \pi0 | 
b^{\dagger}_{\pi0} + b^{\phantom{\dagger}}_{\pi0}
|\text{GS}\rangle
& =
\left(\frac{6 J_{ab} -6 J_{ad} + \Lambda_{y^2} }{6 J_{ac} -6 J_{ad} + \Lambda_{y^2} }\right)^{1/4} .
\end{align}
\end{subequations}
Keeping the leading, physically relevant terms, we get the following magnetic and electric transition matrix elements  
\begin{align}
\langle \pi\pi | M^x | \text{GS} \rangle
&\approx
i \sqrt{3}  g_{xx}  
\left(\frac{ \Lambda_{y^2} }{6 J_{ab} +6 J_{ac} + \Lambda_{y^2} }\right)^{1/4} ,\\
\langle \pi\pi | P^y | \text{GS} \rangle
&\approx
2 i \sqrt{3} c_{2xy} 
\left(\frac{ \Lambda_{y^2} }{6 J_{ab} +6 J_{ac} + \Lambda_{y^2} }\right)^{1/4},
\end{align}
and for the other mode:
\begin{align}
\langle \pi0 |M^x | \text{GS} \rangle
&\approx
\sqrt{3} \left[ g_{xz} 
+  g_{xx} \lambda_{2xz}
\frac{(J_{ac}-J_{ad}) (3 J_{ab}+3 J_{ac}-3 J_{ad}+ \Lambda_{y^2})}{3 ( J_{ab} - J_{ad} ) ( J_{ac} - J_{ad} ) -J_{ad} \Lambda_{y^2}}
 \right] 
\left(
\frac{6 J_{ab} -6 J_{ad} + \Lambda_{y^2} }{6 J_{ac} -6 J_{ad} + \Lambda_{y^2} }
\right)^{1/4}   ,
\\
\langle \pi0 | P^y| \text{GS} \rangle
&\approx
2 \sqrt{3} c_{2yz} 
\left(\frac{6 J_{ab} -6 J_{ad} + \Lambda_{y^2} }{6 J_{ac} -6 J_{ad} + \Lambda_{y^2} }\right)^{1/4} \,.
\end{align}
Using Eq.~(\ref{eq:chiem}), strength of the transition matrix elements of the magnetoelectric susceptibility for the $|\pi\pi\rangle$ mode is: 
\begin{align}
\langle \text{GS}| P^y | \pi\pi  \rangle\langle \pi\pi | M^x | \text{GS} \rangle
&\approx 6
 g_{xx} c_{2xy} 
\left(\frac{ \Lambda_{y^2} }{6 J_{ab} +6 J_{ac} + \Lambda_{y^2} }\right)^{1/2} ,
\end{align}
while for the $|\pi0\rangle$ excitation:
\begin{align}
\langle \text{GS} | P^y| \pi0 \rangle
\langle \pi0 |M^x | \text{GS} \rangle
&\approx
6 c_{2yz} \left[ g_{xz} 
+  g_{xx} \lambda_{2xz}
\frac{(J_{ac}-J_{ad}) (3 J_{ab}+3 J_{ac}-3 J_{ad}+ \Lambda_{y^2})}{3 ( J_{ab} - J_{ad} ) ( J_{ac} - J_{ad} ) -J_{ad} \Lambda_{y^2}}
 \right] 
\left(
\frac{6 J_{ab} -6 J_{ad} + \Lambda_{y^2} }{6 J_{ac} -6 J_{ad} + \Lambda_{y^2} }
\right)^{1/2}   \,.
\end{align}

To summarize, out of the four peaks, two ($|00\rangle$ and $|0\pi\rangle$) are only magnetic dipole active with $\mathbf{H}^\omega\parallel{z}$, while the other two ($|\pi0\rangle$ and $|\pi\pi\rangle$, ME resonances) are both magnetic and electric dipole allowed with $\mathbf{H}^\omega\parallel{x}$ and $\mathbf{E}^\omega\parallel{y}$.
Schematic motion of the local spins (magnetizations) and local polarizations are illustrated in Fig.~\ref{fig:supplfig02} viewed from the $xz$ and $xy$ planes. 
For small values of the single-ion anisotropies $\lambda_{x^2-y^2}$ and $\lambda_{2xz}$, the $\lambda_{x^2-y^2}$  enters into the energy of the modes, splitting the two fold-degenerate modes into four modes, while the $\lambda_{2xz}$ controls the eigenfunctions and therefore the transition matrix elements in the magnetoelectric susceptibility, together with the $c_{2xy}$ and $c_{2yz}$ parameters in the expression for the $P^y$ (\ref{eq:Py}).
The optical directional anisotropy of the two magnetic and electric dipole allowed modes are essentially independent from each other.
Their relative strength, including the sign, is controlled primarily by the ratio of the $c_{2xy}$ and $c_{2yz}$ coefficients in the polarization operator $P^y$, Eq.~(\ref{eq:Py}).
To better understand the role of these parameters we emphasize that the $\mathbf{P}^{\omega}$ oscillating polarization of the ME resonances is built up by the polarization of the $yz$ layers. Phase of the $\mathbf{P}^{\omega}$ compared to the $\mathbf{M}^{\omega}$ sublattice magnetization is affected by the relative phase of the polarizations of these $yz$ layers via the $c_{2xy}/c_{2yz}$ ratio (see Fig.~\ref{fig:supplfig02}(c)).

\begin{figure*}[htbp]
\begin{center}
\includegraphics[width=17.5truecm]{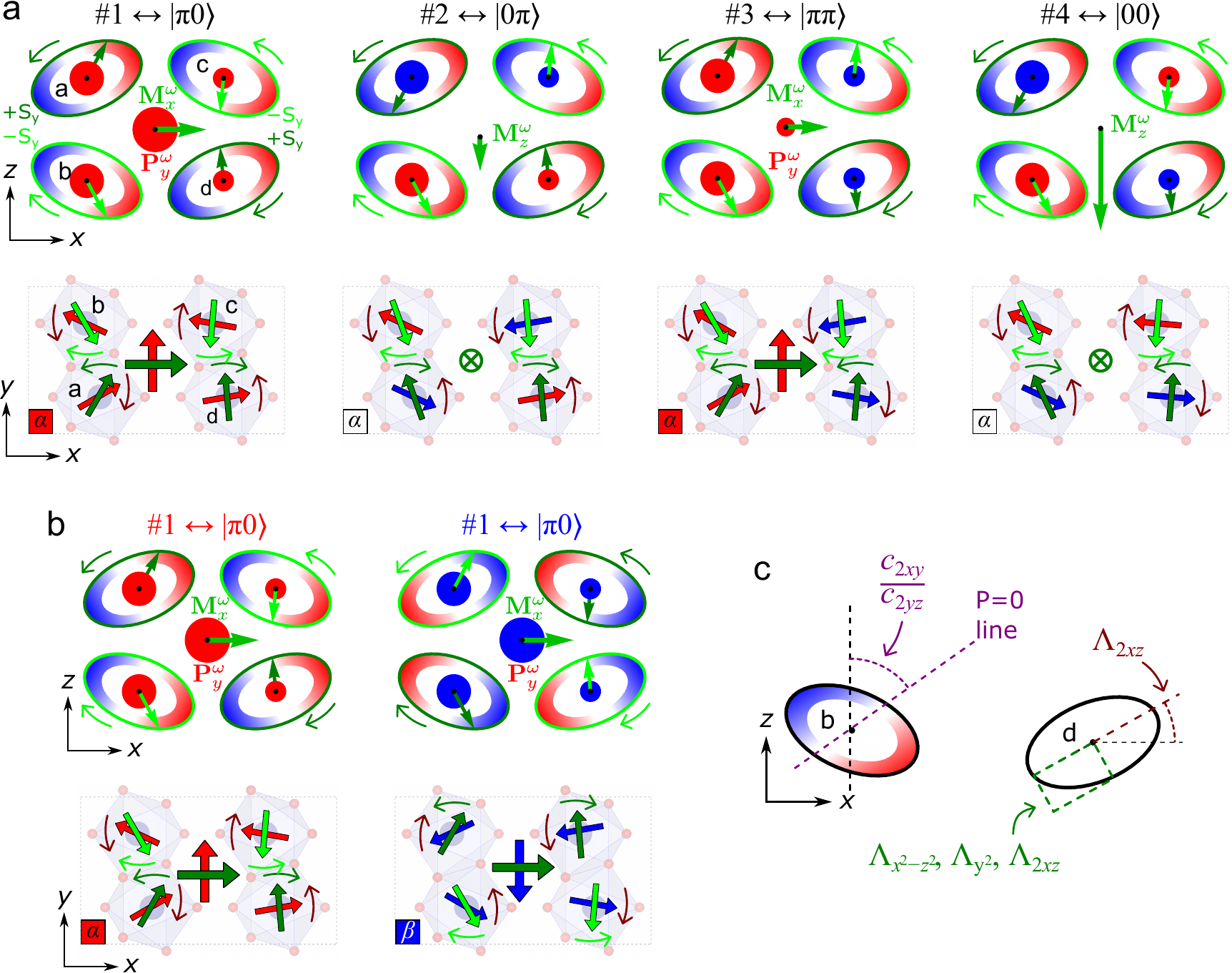}
\caption{\textbf{$\mid$ ME and non-ME resonances of LiCoPO$_4$ viewed from the $xz$ and $xy$ planes.} \textbf{a,} For the sake of simplicity each resonances are illustrated on the $\alpha$ domain for $c_{2xy}>0$. Spins (green and olive arrows) of the ME ($|\pi0\rangle$ and $|\pi\pi\rangle$) and non-ME ($|0\pi\rangle$ and $|00\rangle$) resonances  precesses around canted ellipses in the $xz$ plane. The oscillating $\mathbf{M}^{\omega}$  magnetization and $\mathbf{P}^{\omega}$ polarization of the unit cell are along the $x$ and $y$ axes, respectively, for the ME resonances, while the non-ME resonances have $\mathbf{M}^{\omega}$ along  $z$. While the oscillating polarization (red and blue arrows and dots) of the non-ME resonances are totally canceled out within the $xy$ layers,  the ME resonances have finite $\mathbf{P}^{\omega}$ in the unit cell as a result of the uncompensated polarization of the $xy$ layers. \textbf{b,} The remanent optical ME effect is exemplified on the $|\pi0\rangle$ ME resonance. The ($+E_y^o$,$+H_x^o$) and ($-E_y^o$,$+H_x^o$) poling configurations select the $\alpha$ and $\beta$ domains, respectively. For the same phase of the oscillating magnetization $\mathbf{P}^{\omega}_y$ of the $\alpha$ and $\beta$ domains oscillate in anti-phase with respect to each other, which by means the optical directional anisotropy. \textbf{c,} Instead of circles, the spins precess around ellipses in the $xz$ plane with rotated semi-major axes. Rotation of the semi-major axes depends on the $\Lambda_{2xz}$ parameter while the ellipticity is affected by each on-site anisotropy terms. During the precession of the spin there is an axis across the ellipsis, where $\mathbf{P}^{\omega}=0$. Direction of this line is determined by the ratio of the $c_{2xy}$ and $c_{2yz}$ coefficients.}
\label{fig:supplfig02} 
\end{center}
\end{figure*}


\end{widetext}

\cleardoublepage

\section{Fitting the parameters}

\begin{figure}[htbp]
\begin{center}
\includegraphics[width=8truecm]{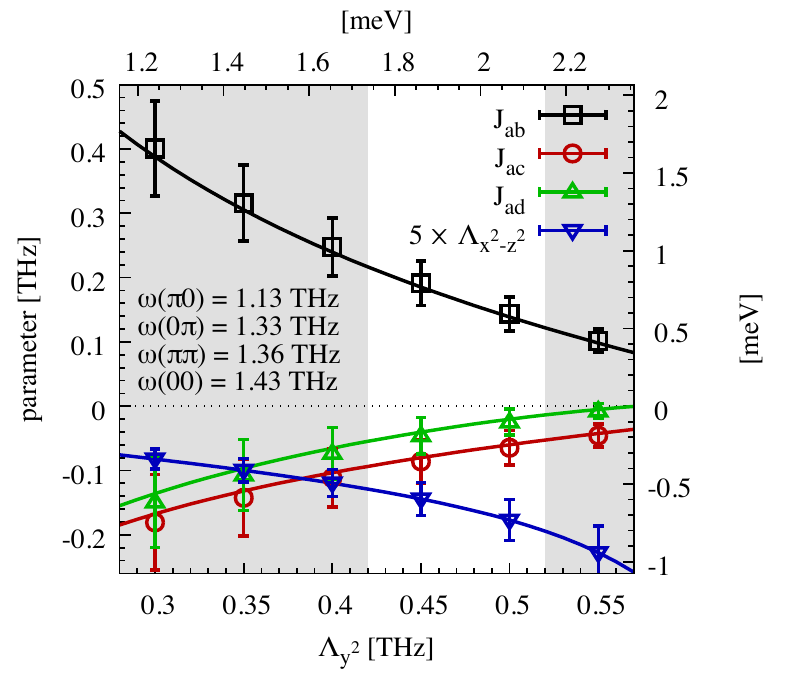}
\caption{\textbf{$\mid$ Fitting of the $J$ exchange and $\Lambda$ single-ion anisotropy parameters.} The exchange couplings $J_{ab}$, $J_{ac}$, and $J_{ad}$, and the anisotropy parameter $\Lambda_{x^2-z^2}$ are determined for fixed values of the $\Lambda_{y^2}$,  assuming the $\#1 \rightarrow \omega_{\pi0}$, $\#2 \rightarrow\omega_{0\pi}$, $\#3\rightarrow\omega_{\pi\pi}$, and $\#4 \rightarrow \omega_{00}$ assignment. As the $\Delta S = 2$ modes are absent from the observed spectral window below 2\,THz, sets an upper limit for the $\Lambda_{y^2}$ at around 0.5\,Thz. The lower limit of 0.42\,THz for $\Lambda_{y^2}$ corresponds to a  3\,THz limit for the energy of the $\Delta S = 2$ modes. This region for the fitting parameters is highlighted by white. The fitting results are in good agreement with the results obtained by neutron scattering measurements~\cite{Tian2008}.}
\label{fig:fitting_main} 
\end{center}
\end{figure}

In the experiment we have identified four modes, which we labeled by numbers form 1 to 4. The peak $\#1$ and $\#3$ show dichroism, therefore they can be assigned to the modes $|\pi0\rangle$ and the $|\pi\pi\rangle$ in some order. Similarly, the remaining peaks $\#2$ and $\#4$ are only magnetically active, so they are assigned to $|00\rangle$ and $|0\pi\rangle$ modes, again, we do not know which one is which. So from the experimental side, we have four input parameters -- the energies of the peaks, and the selection rules restrict the possible number of mode assignments to four. 

  On the theory side, the four input parameter are the $\omega$ energies of the modes (see Eqs.~\ref{eq:omegas}), which depend on five parameters: the three exchange couplings $J_{ab}$, $J_{ac}$, and $J_{ad}$, and the two single-ion anisotropies  $\Lambda_{y^2}$ and $\Lambda_{x^2-z^2}$. The problem is underdetermined at this stage. We have chosen the following strategy to extract the model parameters: we determine the $J_{ab}$, $J_{ac}$,  $J_{ad}$, and $\Lambda_{x^2-z^2}$ by fitting the four experimental energies to $\omega$'s as a function of the $\Lambda_{y^2}$. This has been made for the four possible assignments of the peaks, and we compare them with the existing estimates coming from inelastic neutron scattering measurements~\cite{Tian2008}. We have found that the $(\omega_{\pi0}, \omega_{0\pi}, \omega_{\pi\pi}, \omega_{00})$ order for the peaks $(\#1,\#2,\#3,\#4)$, with energies (1.13 THz, 1.33 THz, 1.36 THz, 1.43THz), is the closest one to the result obtained from the neutrons. The parameter fit as a function of the $\Lambda_{y^2}$ is shown in Fig.~\ref{fig:fitting_main}, and listed for some selected values in Table.~\ref{tab:fitting_main}.
 
To get an estimate of the possible precision of the fitted parameters, we have assumed 10~\text{GHz} standard deviation on the experimental frequencies (corresponding to about 1\% error). The parameters were fitted for 1000 random frequencies with normal distribution with the measured mean value and the assumed standard deviation, the result of this procedure is shown in Fig.~\ref{fig:fitting_main} as error bars. 
Note that the mean values are different from the values calculated exactly at the measured frequencies, as the mean of a nonlinear transformation is not the transformed mean.

To narrow down the possible parameter values shown in Fig.~\ref{fig:fitting_main}, we can use the experimentally observed positions of the $\Delta S=2$ transitions. These modes have so far been omitted from the theoretical discussion, however, we can easily include them. Up to now we only considered the first excited states, $\left|\Psi_{2,i}\right>$ ($i=a,b,c,d$), given by Eq.~(\ref{eq:psi2a}) and Eqs.~(\ref{eq:psi2}). These excitations corresponding to $\Delta S=1$ transitions. The next,  $\Delta S=2$, set of excitations are described by the states  $\left|\Psi_{3,i}\right>$.  $\left|\Psi_{3,a}\right>$ is given by Eq.~(\ref{eq:psi3a}) and we can easily generate the other three wavefunctions corresponding to sublattice $b$, $c$ and $d$:
\begin{subequations}
\begin{eqnarray}
\left|\Psi_{3,b}\right>&=&\left|\uparrow_y\right>+\sqrt{3}\overline{\gamma}\left|\Downarrow_y\right>\\
\left|\Psi_{3,c}\right>&=&\left|\uparrow_y\right>+\sqrt{3}\gamma\left|\Downarrow_y\right>\\
\left|\Psi_{3,d}\right>&=&\left|\downarrow_y\right>+\sqrt{3}\gamma\left|\Uparrow_y\right>
\end{eqnarray}
\end{subequations}
We introduce bosons that create these $\Delta S=2$ excitation with the following notation; $c^\dagger_i\left|0\right>=\left|\Psi_{3,i}\right>$, where the vacuum state $\left|0\right>$ corresponds to the ground state $\prod_n\left|\Psi_{1,i}\right>$, i.e. the vacuum of excitations.

The Hamiltonian for the bosons $c^\dagger_i$ is already diagonal and, as it turns out, the four modes are degenerate in zero fields, so
\begin{align}
 \mathcal{H}_{\Delta S=2} = 
\omega_c \sum_{i=a,b,c,d} c^{\dagger}_{i} c^{\phantom{\dagger}}_{i}
\end{align}
with the excitation energy 
\begin{align}
\omega_c=12 J_{ab} + 12 J_{ac} - 12 J_{ad} + 2 \Lambda_{y^2}\;.
\end{align}

From the absorption spectra, below 2\,THz we do not see additional modes to the four $\Delta S=1$ excitations, excited by $b^\dagger_{00}$, $b^\dagger_{0\pi}$, $b^\dagger_{\pi0}$ and $b^\dagger_{\pi\pi}$. The new excitations are expected between 2\,THz and 3\,THz. Thus, $\omega_c$ needs to be in this regime, allowing us to constrict the coupling parameters. The invalid parameter region $\omega_c \alt 2$\,THz corresponds to $\Lambda_{y^2}\agt$0.52\,THz, shown as the gray area above 0.52\,THz in Fig.~\ref{fig:fitting_main}. While the $\omega_c \agt 3$\,THz  region belongs to the gray sector below  $\Lambda_{y^2} \approx 0.42$\,THz in Fig.~\ref{fig:fitting_main}, setting the lower boundary for $\Lambda_{y^2}$. The white region in Fig.~\ref{fig:fitting_main} illustrates the expected valid parameter range.

\begin{figure}[htbp]
\begin{center}
\includegraphics[width=8truecm]{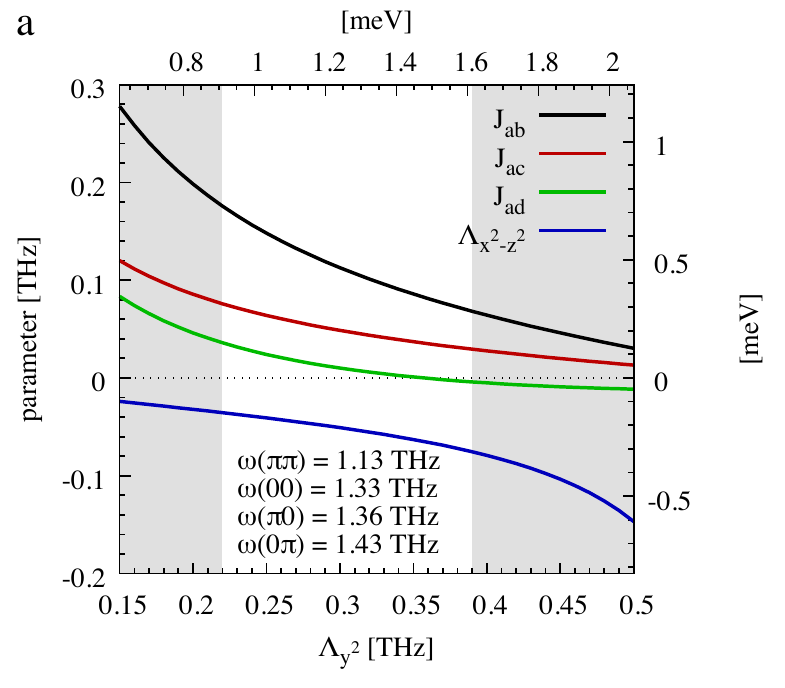}
\includegraphics[width=8truecm]{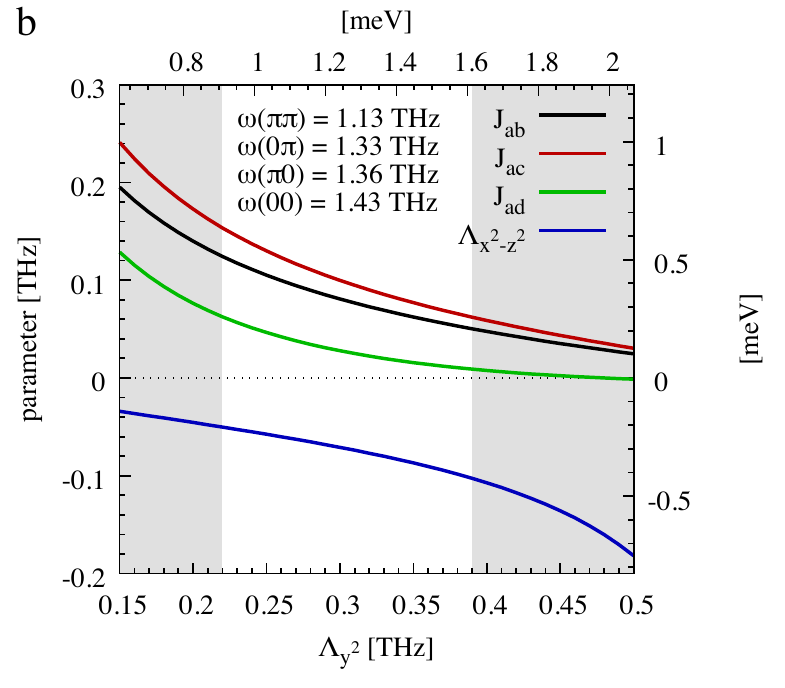}
\caption{\textbf{$\mid$ Exchange and single-ion anisotropy parameters for different assignment of the peaks.}
\textbf{a,} Relationship between the fitting parameters for $(\#1,\#2,\#3,\#4) \rightarrow (\omega_{\pi\pi},\omega_{00},\omega_{\pi0},\omega_{0\pi})$ assignment, and \textbf{b,} for $(\#1,\#2,\#3,\#4) \rightarrow (\omega_{\pi\pi},\omega_{0\pi},\omega_{\pi0},\omega_{00})$ assignment of the observed magnetic resonances. In these cases, the fitted parameters show significant difference from the results of the neutron diffraction.}
\label{fig:fitting_alt1} 
\end{center}
\end{figure}

 The exchange parameters for the other assignments, shown in Figs.~\ref{fig:fitting_alt1}, are less likely as the signs of the exchange couplings are in contradiction with the corresponding parameters from the neutron study. The fourth assignment, not shown, gives values with even larger difference.

\begin{table}[t]
\begin{tabular}{ccccc}
$\Lambda_{y^2}$ & $J_{ab}$ & $J_{ac}$ & $J_{ad}$ & $\Lambda_{x^2-z^2}$
 \\
\hline
\hline
0.5 & 0.143 & -0.065 & -0.024 & -0.036 \\
0.45 & 0.191 & -0.085 & -0.045 & -0.029 \\
0.4 & 0.248 & -0.112 & -0.074 & -0.024\\
\hline
\end{tabular}
\caption{\textbf{$\mid$ The fitted exchange and single-ion anisotropy parameters.} for different values of the $\Lambda_{y^2}$ parameter, assuming the $\#1 \rightarrow \omega_{\pi0}$, $\#2 \rightarrow\omega_{0\pi}$, $\#3\rightarrow\omega_{\pi\pi}$, and $\#4 \rightarrow \omega_{00}$ assignment, the same  as in Fig.~\ref{fig:fitting_main}. All parameters are shown in THz unit.
\label{tab:fitting_main}
}
\end{table}

\begin{references}

\bibitem{Fiebig2016}
Fiebig, M., Lottermoser, T., Meier, D. \& Trassin, M.
The evolution of multiferroics. {\it Nat. Rev. Mats.}
{\bf 1}, 16046 (2016).

\bibitem{Kimura2007}
Kimura, T. {\it et al.}
Magnetic control of ferroelectric polarization. {\it Nature} {\bf 426}, 55-58 (2003).

\bibitem{Dong2015}
Dong, S., Liu, J-M., Cheong, S-W. \& Ren, Z.
Multiferroic materials and magnetoelectric physics: symmetry, entanglement, excitation, and topology.
{\it Advances in Physics} {\bf 64}, 519-626 (2015).

\bibitem{Fiebig2005}
Fiebig, M. Revival of the magnetoelectric effect. {\it J. Phys. D.: Appl. Phys.} {\bf 38}, R123-R152 (2005).

\bibitem{Fiebig2005_2}Spaldin, N. A. \& Fiebig, M. The Renaissance of Magnetoelectric Multiferroics. {\it Science} {\bf 309}, 391-392 (2005).

\bibitem{Eerenstein2006} Eerenstein, W., Mathur, N. D. \& Scott, J. F. Multiferroic and magnetoelectric materials. {\it Nature} {\bf 44}, 759-765 (2006).

\bibitem{Cheong2007}Cheong, S.-W. \& Mostovoy, M. Multiferroics: a magnetic twist for ferroelectricity. {\it Nat. Mater.} {\bf 6}, 13-20 (2007).

\bibitem{Sando2013}Sando, D. {\it et al.} Crafting the magnonic and spintronic response of
BiFeO$_3$ films by epitaxial strain. {\it Nat. Mater.} {\bf 12},
641-646 (2013).

\bibitem{Henron2014}Henron, J. T. {\it et al.} Deterministic switching of ferromagnetism at room temperature using an electric field. {\it Nature} {\bf 516}, 370-373 (2014).

\bibitem{Kezsmarki2015} K\'ezsm\'arki, I. {\it et al.} Optical diode effect at spin-wave excitations of the room-temperature multiferroic BiFeO$_3$. {\it Phys. Rev. Lett.} {\bf 115}, 127203 (2015).

\bibitem{Mercier1967} Mercier, M., Gareyte, J. \& Bertaut, E. F. Une nouvelle famille de corps magnetoelectrique -- LiMPO$_4$ (M= Mn, Co, Ni). {\it C. R. Seances Acad. Sci., Ser. B} {\bf 264}, 979 (1967).

\bibitem{Rivera1994} Rivera, J. P. The linear magnetoelectric effect in LiCoPO$_4$
revisited. {\it Ferroelectrics} {\bf 161}, 147-164 (1994).

\bibitem{VanAken2007}Van Aken, B.B., Rivera, J-P., Schmid, H. \&
Fiebig, M. Observation of ferrotoroidic domains. {\it Nature} {\bf
449}, 702-705 (2007).

\bibitem{Zimmermann2014}
Zimmermann, A. S., Meier, D. \& Fiebig, M.
Ferroic nature of magnetic toroidal order.
{\it Nat. Commun.} {\bf 5}, 4796 (2014).

\bibitem{Jungwirth2016}
Jungwirth, T. and Marti, X. and Wadley, P. and Wunderlich, J.
Antiferromagnetic spintronics
{\it Nat. Nano.} {\bf 11}, 231-241 (2016).

\bibitem{Santoro1966} Santoro, R. P., Segal, D. J. \& Newnham, R. E. Magnetic properties of LiCoPO$_4$ and LiNiPO$_4$. {\it J. Phys.
Chem. Solids} {\bf 27}, 1192-1193 (1966).

\bibitem{Kezsmarki2011} K\'ezsm\'arki, I. {\it et al.} Enhanced directional dichroism of terahertz light in resonance with magnetic excitations of the multiferroic Ba$_2$CoGe$_2$O$_7$ oxide compound.
{\it Phys. Rev. Lett.} {\bf 106}, 057403 (2011).

\bibitem{Miyahara2012} Miyahara, S. \& Furukawa, N. Nonreciprocal directional dichroism and toroidalmagnons in helical magnets.
{\it J. Phys. Soc. Jpn.} {\bf 81}, 023712 (2012).

\bibitem{Kezsmarki2014} K\'ezsm\'arki, I. {\it et al.} One-way transparency of four-coloured spin-wave excitations in multiferroic materials. {\it Nat. Commun.} {\bf 5}, 3203 (2014).

\bibitem{Bordacs2012}Bord\'acs, S. {\it et al.} Chirality of matter shows up via spin excitations. {\it Nat. Phys.} {\bf 8}, 734-738 (2012).

\bibitem{Arima2008}Saito, M., Ishikawa, K., Taniguchi, K. \& Arima, T. Magnetic control of crystal chirality and the existence of a large magneto-optical dichroism effect in CuB$_2$O$_4$.
{\it Phys. Rev. Lett.} {\bf 101}, 117402 (2008).

\bibitem{Takahashi2012}Takahashi, Y., Shimano, R., Kaneko, Y., Murakawa, H. \& Tokura, Y. Magnetoelectric resonance with electromagnons in a perovskite helimagnet.
{\it Nat. Phys.} {\bf 8}, 121-125 (2012).

\bibitem{Penc2012}
Penc, K., Romh\'anyi, J., R{\~o}{\~o}m, T., Nagel, U., Antal, \'A., Feh\'er, T., J\'anossy, A., Engelkamp, H. , Murakawa, H., Tokura, Y., Szaller, D., Bord\'acs, S. \& K\'ezsm\'arki, I.
Spin-Stretching Modes in Anisotropic Magnets: Spin-Wave Excitations in the Multiferroic Ba$_2$CoGe$_2$O$_7$
{\it Phys. Rev. Lett.} {\bf 108}, 257203 (2012).

\bibitem{Miyahara2011}
Miyahara, S., \& Furukawa, N. Theory of Magnetoelectric Resonance in
Two-Dimensional $S$ = 3/2 Antiferromagnet Ba$_2$CoGe$_2$O$_7$ via
Spin-Dependent Metal-Ligand Hybridization Mechanism. {\it J. Phys.
Soc. Jpn.} {\bf 80}, 073708 (2011).

\bibitem{Tian2008}
Tian, Wei, Li, Jiying, Lynn, J. W., Zarestky, J. L., \& Vaknin, D.
Spin dynamics in the magnetoelectric effect compound LiCoPO$_4$.
{\it Phys. Rev. B} {\bf 78}, 184429 (2008).

\bibitem{Arima2007}
Arima T., Ferroelectricity Induced by Proper-Screw Type Magnetic
Order. {\it J. Phys. Soc. Jpn.} {\bf 76}, 073702 (2007).

\bibitem{Vieweg2014} Vieweg, N., Rettich, F., Deninger, A., Roehle, H., Dietz, R., \&
G\:obel, T. A time-domain terahertz spectrometer with 90 dB dynamic
range. {\it J. Infrared Millim. Terahertz Waves} {\bf 35}, 823
(2014).

\bibitem{SaintMartin2008} 
Saint-Martin, R. \& Sylvain Franger, S.,
Growth of LiCoPO$_4$ single crystals using an optical floating-zone technique
{\it J. Cryst. Growth} {\bf 310}, 861-864 (2008).


\bibitem{Romhanyi2011}
Romh\'anyi, J., Lajk\'o, M., \& Penc, K.
Zero- and finite-temperature mean field study of magnetic field induced electric polarization in Ba$_2$CoGe$_2$O$_7$: Effect of the antiferroelectric coupling
{\it Phys. Rev. Lett.} {\bf 84}, 224419 (2011).

\bibitem{Romhanyi2012}
Romh\'anyi, J., \& Penc, K.
Multiboson spin-wave theory for Ba$_2$CoGe$_2$O$_7$: A spin-3/2 easy-plane N\'eel antiferromagnet with strong single-ion anisotropy
{\it Phys. Rev. B} {\bf 86}, 174428 (2012).

\end{references}
\end{document}